\documentclass[twocolumn, preprint]{aastex631}
\usepackage{graphicx} 
\usepackage{amsmath}
\usepackage{appendix}

\usepackage{xcolor}

\usepackage[commentmarkup=uwave,commandnameprefix=ifneeded]{changes}
\renewcommand{\replaced}{\chreplaced}
\renewcommand{\deleted}{\chdeleted}

\newcommand{\Mtr}{M_{\mathrm{tr}}}
\newcommand{\Mtrdot}{\dot{M}_{\mathrm{tr}}}
\newcommand{\MI}{M_{\mathrm{I}}}
\newcommand{\MBHMax}{M_{\mathrm{BHmax}}}

\definechangesauthor[name=Max, color=teal]{MI}
\definechangesauthor[name=Jacob, color=orange]{JG}
\definechangesauthor[name=Will, color=magenta]{WF}

\newcommand{\CCA}{\affiliation{Center for Computational Astrophysics, Flatiron Institute, 162 5th Ave, New York, NY 10010, USA}}
\newcommand{\CIT}{\affiliation{Department of Physics, California Institute of Technology, Pasadena, California 91125, USA}}
\newcommand{\CITLab}{\affiliation{LIGO Laboratory, California Institute of Technology, Pasadena, California 91125, USA}}
\newcommand{\StonyBrook}{\affiliation{Department of Physics and Astronomy, Stony Brook University, Stony Brook NY 11794, USA}}

\begin{document}
\title{Physical Models for the Astrophysical Population of Black Holes: Application to the Bump in the Mass Distribution of Gravitational Wave Sources}

\author{Jacob Golomb}
\email{jgolomb@caltech.edu}
\CIT
\CITLab
\author{Maximiliano Isi}
\email{misi@flatironinstitute.org}
\CCA
\author{Will M. Farr}
\email{wfarr@flatironinstitute.org}
\CCA
\StonyBrook
\date{\today}

\begin{abstract}
Gravitational wave observations of binary black holes have revealed unexpected
structure in the black hole mass distribution. Previous studies employ
physically-motivated phenomenological models and infer the parameters that
control the features of the mass distribution that are allowed in their model,
associating the constraints on those parameters with their physical motivations
\textit{a posteriori}. In this work, we take an alternative approach in which we
introduce a model parameterizing the underlying stellar and core-collapse
physics and obtaining the remnant black hole distribution as a derived
byproduct. In doing so, we constrain the stellar physics necessary to explain
the astrophysical distribution of black hole properties under a given model. We
apply this to the mapping between initial mass and remnant black hole mass,
accounting for mass-dependent mass loss using a simple parameterized
description.  Allowing the parameters of the initial mass-remnant mass
relationship to evolve with redshift permits correlated and physically
reasonable changes to features in the mass function. We find that the current
data are consistent with no redshift evolution in the core-remnant mass
relationship, but place only weak constraints on the change of these parameters.
This procedure can be applied to modeling any physical process underlying the
astrophysical distribution.  We illustrate this by applying our model to the
pulsational pair instability supernova (PPISN) process, previously proposed as
an explanation for the observed excess of black holes at ${\sim} 35\, M_\odot$.
Placing constraints on the reaction rates necessary to explain the PPISN
parameters, we concur with previous results in the literature that the peak
observed at ${\sim} 35\, M_\odot$ is unlikely to be a signature from the PPISN
process as presently understood.  
\end{abstract}

\section{Introduction}

Observations of gravitational waves from binary-black-hole (BBH) and binary-neutron-star (BNS) mergers with the LIGO-Virgo-KAGRA detector network have provided otherwise inaccessible information on the properties of those compact objects \citep{LVK, LIGO, Virgo}. While individual events offer a glimpse into the details of a particular black hole (BH) or neutron star, studying observations collectively  on a population-level allows us to draw inferences about stellar astrophysics, the formation channels of BHs and neutron stars, the overall rates at which BBH/BNS mergers occur in the universe, and cosmology \citep{gwtc3pops, gwtc3cosmo, gwtc3, gwtc2pops}, as well as tests of general relativity \citep{LIGOScientific:2021sio,Payne:2023kwj}.

Models adopted for population inference of BBHs tend to take one of two major approaches. The first are so-called parametric methods, in which a phenomenological model is constructed using relatively few parameters, with these parameters directly controlling well-defined features encoded in the model. This commonly involves assuming a functional form for the global structure of the distribution (e.g., a truncated power law for the mass distribution), enhanced by features such as a bump, dip, or break, and jointly fitting for the properties of the global structure and additional features \citep{TalbotThranePLP, gwtc2pops, gwtc3pops, Kovetz17, Fishbach17}. The other popular approach consists of data-driven methods (sometimes called ``non-parametric" methods, in spite of their overabundance of parameters) in which a flexible model is allowed to fit nearly-arbitrary shapes to the distribution. Such fits have been achieved with tools such as splines, histograms, and Gaussian process regression \citep{Callister23, Golomb22, Mandel18, Ray23, Edelman22, Tiwari21}.

While the latter method is more general and can capture features not explicitly defined in a model, the former offers the ability to encode signatures from expected physical processes in parameters controlling the shape of the distribution, making it possible to interpret the constraints in terms of the underlying physical motivations. Nevertheless, interpreting the constraints on these parameters as constraints on the underlying physics can be difficult, as there are often unmodeled assumptions as to how the underlying physical processes translate into resulting distribution that is being modeled.

In this work, we provide an alternative approach to prescribing a parametric population model: instead of modeling the BBH distribution directly, we introduce parameters to describe the underlying astrophysics and the associated mapping to remnant BH properties and we derive the resulting population distribution as a result of these underlying parameters and its associated mapping. Subject to model assumptions, we more directly infer the physics as informed by BBH properties in a way that avoids strong phenomenological approximations in the BBH population model itself.
This is related to the ``backpropagation" approach in \cite{Wong22} and the method in \cite{Andrews21}, except we operate at the population level rather than individual events and impose a rigid physical mapping from progenitors to remnants.

Previous parametric approaches to modeling the BBH mass distribution have been particularly useful to place constraints with relatively little data when strong assumptions about the structure of the mass distribution are warranted. For example, parametric population analyses of the first catalogs of BBH events have revealed that the mass distribution is well-described by a truncated power law that peaks at ${\sim} 8\, M_\odot$, decaying to high masses, and featuring an overdensity (modeled as a Gaussian bump) at ${\sim}\, 35\, M_\odot$ \citep{gwtc3pops, gwtc2pops}.

We apply our approach to the overdensity in the mass distribution at ${\sim} 35\, M_\odot$, which may or may not be accompanied by a subsequent dip \citep{TalbotThranePLP, Edelman22}. The original motivation for looking for this feature was the expected ``pile-up" of BHs that resulted from progenitors that had undergone pair instability pulsations \citep{TalbotThranePLP}.
This pulsational-pair instability supernova (PPISN) process results in a nonlinear mapping from the masses of progenitors stars to their remnant BHs \citep{Woosley17, Woosley19, Woosley21, Farmer19}: for relatively low masses, the mapping is linear; however, it turns approximately quadratic for higher masses, with a turnover that caps the a maximum BH mass obtainable from the evolution of an individual star; in BH-mass space, the beginning of this BH ``mass gap'' is preceded by a pile-up from the quadratic turnover.
This kind of relationship between progenitor and BH masses can be directly exploited to bridge parametric models of BBH populations with stellar physics for improved inference, as we do here.

In this work, we implement a simple model for the initial mass function of
stellar progenitors and the associated map from progenitor mass to remnant mass,
motivated by the type of relationship found from the simulations in
\cite{Farmer19, Farmer20}.  Rather than informing our model with individual massive
sources, we construct a full population model for the mass distribution,
including a subpopulation in the upper mass gap due to higher-generation
mergers. Using data from gravitational wave events from the third Gravitational
Wave Transient Catalog \citep[GWTC-3,][]{gwtc3}, we infer the shape of the
initial mass function stellar progenitors, the associated mapping to the remnant
BH distribution, and the relative contribution of sources formed through 1G
mergers. 

\citet{Baxter21} also explored the consequences of a similar initial
mass-remnant mass relation to those in \citet{Farmer19,Farmer20}, but unlike our
work here that directly parameterizes the relation, they used parameterized
models that had been fitted to theoretical BH mass functions.

Having a physically-motivated model for the BH mass distribution
facilitates extensions that incorporate richer physics.  As an example, here we
allow the underlying physics to evolve with redshift, as may be expected from
cosmic history considerations. Such evolution in the underlying physical
parameters captures the correlated changes in shape and height of the bump that
must occur in the presence of changing progenitor to remnant mass relationships.
It may be possible to use shape measurements to calibrate changes in the mass
scale of the bump with redshift to reduce or eliminate systematic uncertianities
in cosmological parameter inference from the BBH mass function, sometimes called
the ``spectral siren'' method \citep{Farr19b,Ezquiaga2022}.

While here we apply this approach to modeling the PPISN process underlying the astrophysical BBH mass distribution, it can more generally be used as a model to place constraints on any relationship between progenitor mass and remnant BH mass as informed by gravitational wave observations. The model introduced here can readily be applied to any process with accelerating mass loss as a function of progenitor mass, but this method can be useful for inferring the physics of any arbitrary relationship underlying an observable distribution associated with BBHs \citep[cf.][for a related approach applied to inferring the delay times between binary formation and merger]{Fishbach23}.

We begin with an overview of hierarchical Bayesian inference in Sec.~\ref{sec:hierarchicalinference}. We then outline the models with and without evolution with redshift in Sec.~\ref{sec:models}. In Sec.~\ref{sec:results}, we present results for both model configurations, using data from the third oberseving run (O3) of LIGO-Virgo. We offer interpretations of our results in Sec.~\ref{sec:discussion} and provide concluding remarks in Sec.~\ref{sec:conclusions}. We find that the PPISN mechanism, as currently predicted by stellar evolution models, cannot predict the 35 $M_\odot$ feature in the BBH mass distribution.

\section{Hierarchical Bayesian Inference}\label{sec:hierarchicalinference}

We conduct our inference on the population parameters $\Lambda$ with a hierarchical Bayesian framework, in which we inform our population model with a catalog $N_{\rm det}$ events, to compute the likelihood (see, e.g., \cite{Mandel19, Thrane19}):
\begin{equation}\label{likelihood}
    \mathcal{L}(\{d\}|\Lambda) \propto \frac{K(\Lambda)^{N_d} e^{-K(\Lambda)}}{p_{\rm det}(\Lambda)^{N_d}} \prod_{i=1}^{N_{\rm det}} \int\mathcal{L}(d_i|\theta) \pi(\theta|\Lambda)d\theta
\end{equation}
where $\mathcal{L}(d_i|\theta)$ is the likelihood of the data for the $i$th event, given physical parameters $\theta$ (i.e., masses, distances), and $\pi(\theta|\Lambda)$ is our population model with a predicted number of detections $K$. The $p_{\rm det}(\Lambda)$ prefactor accounts for the selection effects associated with observing a catalog biased toward sources with parameters that favor detectability (i.e., the Malmquist bias). See Appendix \ref{app:derivation} for details on this likelihood.

Following the approach in \cite{Farr19} and \cite{Tiwari18}, we compute $p_{\rm det}(\Lambda)$ with injections of sources from a fiducial population in detector noise, and assigning weights to each of the sources that pass our detection threshold. These sensitivity injections are from the O3 injection set released in \cite{O3injections}. We compute the per-event population evidence (the integral in Eq.~\eqref{likelihood}) by reweighting samples from individual event posterior distributions and dividing by the event-specific sampling priors. Since our population model is written only in terms of masses and distances, we effectively adopt the prior from parameter estimation for the spin parameters (isotropic in direction and uniform in spin magnitude).

For our analyses involving third observing run (O3) data, we obtain posterior samples for each event from \cite{gwtc3PEdata}, using the same BBH events from O3 as in \cite{gwtc3pops}. This results in 59 events meeting the False Alarm Rate (FAR) threshold of 1 per year.  Throughout this work we assume the best-fit cosmological parameters from the Planck 2018 release \citep{Planck18}.

We sample the population posterior using the No-U-Turn-Sampler (NUTS) in \texttt{Numpyro} \citep{phan19, bingham19}, and we write the functions for computing Eq.~\eqref{likelihood} in \texttt{jax} \citep{jax2018github} to take advantage of automatic differentiation when sampling with Hamiltonian Monte Carlo \citep{Duane87}.

 We do not enforce the convergence conditions from \cite{gwtc3pops} for the Monte Carlo integrals in our likelihood, but we confirm that all points in our posterior have a reasonable enough number of effective per event posterior samples and injections for good convergence.

\section{Mass Distribution of Black Holes From Progenitor Mass Function}\label{sec:models}

\subsection{Mass Distribution Model}\label{Mass}

We begin to construct our mass distribution model by assuming a functional form
for the initial mass function (IMF) of compact object progenitors. Surveys have
shown that the stellar IMF on the main sequence can be well-modeled as a
featureless power law at high masses, with a power law index of approximately
$-2.3$ \citep[see, e.g.,][]{Kroupa01, Salpeter55, Kroupa19}. Recent studies have
shown through simulations and analytic approximations that the there may be an
approximately linear relationship between a high-mass star's zero-age
main-sequence (ZAMS) mass and the mass of its core before undergoing supernova,
although this is uncertain \citep{Woosley19, Belczynski16}. We therefore assume
that the IMF of compact object progenitors can also be modeled with a power law
to good approximation, but allow for a break at $20\, M_\odot$ for additional
flexibility \citep[see, e.g.,][]{Schneider18}. Even if this relationship has
nontrivial nonlinearities, modelling the shape of the broken power law should
capture the dominant resolvable structure of the distribution.

We express the distribution of initial progenitor masses, $\MI$, as
\begin{equation}\label{dNdMCO}
    \frac{dN}{d\MI}(\MI) \propto
    \begin{cases}
        \left(\frac{\MI}{20 M_\odot}\right)^{-a} & \text{if } \MI < 20 M_\odot \, , \\
        \left(\frac{\MI}{20 M_\odot}\right)^{-b} & \text{if } \MI > 20 M_\odot \, .
    \end{cases}
\end{equation}

In order to obtain the resulting BH mass distribution from the progenitor mass
distribution, we require a mapping between $\MI$ for a progenitor and the mass
of its remnant after undergoing core collapse.  Here we assume that the mean
remnant mass follows the initial mass for small initial masses, before smoothly
transitioning at black hole masses $\Mtr$ to a quadratic relationship that
exhibits a maximum remnant mass $\MBHMax$, eventually decaying to zero remnant
masses.  We impose throughout this work a constraint that $\MBHMax > \Mtr$ so
that our mapping is well-defined.  We express this piecewise mapping through a
functional form $\bar{M}_{\rm BH}(\MI|\Mtr, \MBHMax)$ given by
\begin{widetext}
\begin{equation}\label{mumbh}
    \bar{M}_{\rm BH} (\MI|\Mtr, \MBHMax) = 
    \begin{cases}
        \MI & \text{if } \MI < \Mtr\, , \\
        \MBHMax + \frac{(\MI - 2 \MBHMax + \Mtr)^2}{4 (\Mtr - \MBHMax)} & \text{if } \Mtr < \MI < 2 \MBHMax - \Mtr\, ,\\
        0 & \text{otherwise}.
    \end{cases}
\end{equation}
\end{widetext}

Such a simple model will inevitably miss some of the complexity of the
initial-final mass relationship.  We introduce scatter in the remnant mass at
fixed initial mass to account for this missing physics.  We simulate such
uncertainty in the $\MI-M_{\rm BH}$ mapping by treating the natural logarithm of
the remnant mass as a realization from a Gaussian distribution, with standard
deviation $\sigma$:
\begin{multline}\label{pmbh}
    p\left(\ln(M_{\rm BH}) \mid \bar{M}_{\rm BH}, \sigma \right) = \\ \mathcal{N}\left[\ln\left(\bar{M}_{\rm BH}(\MI)\right), \sigma\right]\left( \ln(M_{\rm BH}) \right) \, .
\end{multline}
where $\bar{M}_{\rm BH}(M_{\rm I})$ is given in Eq.~\eqref{mumbh}. Since
Eq.~\eqref{pmbh} specifies that the logarithm of $M_{\rm BH}$ values are
normally distributed around with standard deviation $\sigma$, the
uncertainty on the physical value of the mass $M_{\rm BH}$ will grow with $\MI$
for fixed $\sigma$.  

Any confident measurement of a nonzero value of $\sigma$ would mean there is
variation in the $\MI-M_{\rm BH}$ mapping. This could originate from any number
of factors, e.g.\ physical properties affecting stellar evolution manifesting
differently between black holes in the catalog. For example, since metallicity
is expected to have a slight effect on the remnant mass given an initial CO core
mass in models of the pair instability \citep{Farmer19}, resolvable
contributions from sources with differing birth metallicities in our dataset
would result in a preferentially nonzero value for $\sigma$.

\begin{figure}
    \centering
    \includegraphics[scale=0.4]{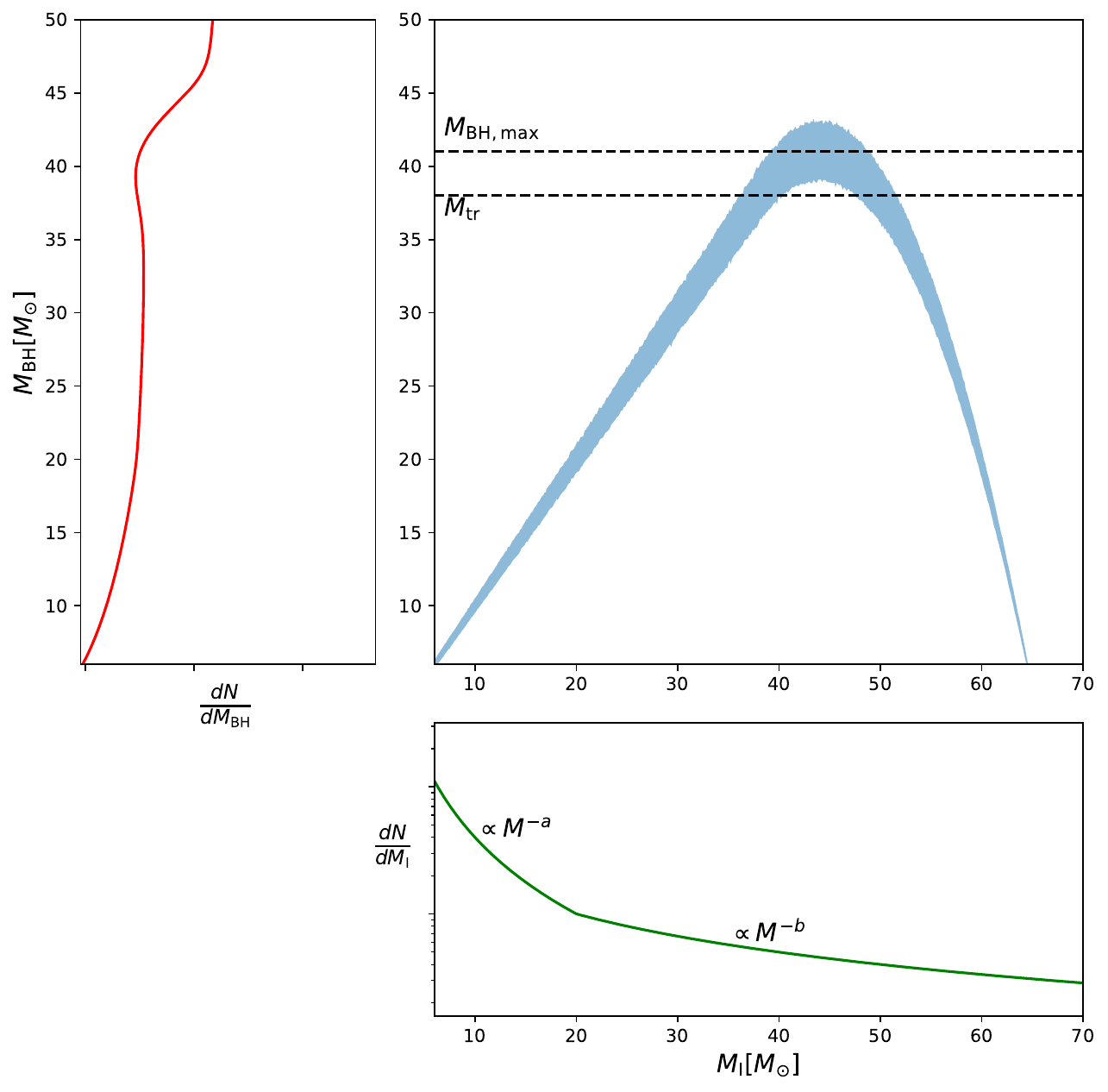}
    \caption{Obtaining a 1G BH mass distribution from the initial mass function.
    The progenitor IMF (bottom panel) gets transformed through the $\MI$ to
    $M_{\rm BH}$ mapping (top right panel), resulting in a distribution for
    ${dN}/{dM_{1G}}$ of BHs in 1G systems. We label parameters underlying the 1G
    BBH mass distribution: $a$ and $b$ are the low-mass and high-mass spectral
    indices of the progenitor IMF; $\Mtr$ and $M_{\rm BH,max}$ (dashed lines)
    control the onset of the nonlinearity and the maximum of the quadratic part
    of the mapping, respectively. The blue shaded region is the 90\% credible
    region of the lognormal mapping for our choice of $\sigma$ (see
    Eq~\ref{dNdM1g} and preceding equations for the functional form). We vary
    these parameters in our fit to the LIGO-Virgo data, together with the
    parameters for the 2G population (see Section
    \ref{sec:nonevolvingmassmodel}).}
    \label{fig:imf_mass_spectrum}
\end{figure}

To obtain the mass distribution for stellar-origin (``first generation'') BHs
${dN}/{dM_{1G}}$, we integrate over progenitor masses,
\begin{equation}\label{dNdM1g}
\begin{split}
    \frac{dN}{dM_{\rm 1G}}& \left(M_{\rm BH} \mid a, b, \Mtr, M_{\rm BH,max}, \sigma \right) = \\ & \int d\MI \frac{dN}{d\MI} \, p(M_{\rm BH} \mid \MI).
\end{split}
\end{equation}
(Note the implicit Jacobian from the logarithmic mass appearing in
Eq.~\eqref{pmbh}.)  The turnover in the $\MI - M_{\rm BH}$ relation above $\Mtr$
leads to a pile-up of black hole masses around $M_{\rm BH,max}$.  This pile-up
is usually expressed as a relative overdensity through a Gaussian bump
\citep{TalbotThranePLP, gwtc2pops, gwtc3pops}; in our model, the location,
width, height, and asymmetry of the bump are ``naturally'' derived from the
parameters $\Mtr$, $\MBHMax$, and $\sigma$.  The black hole mass functions in
our model (see Figure \ref{fig:imf_mass_spectrum}) are similar to those
discussed in \citet{Baxter21} (they would agree in the limit of $\sigma \to 0$),
but those authors did not attempt to describe the black hole mass function in
terms of the underlying physical processes relating initial to final mass as we
do here.

This simple, general parameterized model is deliberately reminiscent of the
relationship expected between CO core mass and remnant BH mass induced by the
pair instability \citep{Fowler64,Rakavy67}.  From simulations with the stellar
evolution code \texttt{MESA} \citep{Paxton19}, \cite{Farmer19} find that $\MI$
prior to core collapse is in fact the dominant variable determining the remnant
mass post core collapse. Figure 4 in \cite{Farmer19} shows the resulting $M_{\rm
BH}$ vs $\MI$ relationships obtained for a range of choices of input physics and
metallicity. The authors note that for a given choice of metallicity, this
relationship is well-modeled by a piecewise map: a linear relationship, turning
over to a quadratic at the CO core mass at which pulsations begin to remove
notable mass, followed by a decay to $M_{\rm BH} = 0\, M_\odot$, corresponding
to the mass at which pulsational pair instability fully disrupts the star,
leaving no remnant. This general trend has been confirmed by other
simulation-based studies \citep[e.g.,][]{Mehta22, Woosley17}.

When $\MI$ reaches $\Mtr$, the $\MI$ to $M_{\rm BH}$ mapping transitions from
its linear relationship to a nonlinear one.  In terms of the pair instability,
at this point the pulsation process causes mass loss whose efficiency increases
with the star's mass \citep{Marchant19, Farmer19, Woosley17, Woosley21}. The
form of the quadratic function in Eq.~\eqref{mumbh} puts the peak value at
$M_{\rm BH,max}$ and enforces that the transition and its derivative be
continuous at $\Mtr$. This results in a BH mass distribution in which remnants
between roughly $\Mtr$ and $M_{\rm BH,max}$ can map back to a wider range of
progenitor masses, and each BH mass bin $dM_{\rm BH}$ in this range contains
more systems than it otherwise would had the $\MI - M_{\rm BH}$ relationship
continued to be linear.

As shown in \cite{Farmer19}, the map from $\MI$ to $M_{\rm BH}$ is sensitive to
unknown physics affecting the core collapse and stellar evolution process, even
given a fixed $\Mtr$. We therefore do not know with certainty of a one-to-one
map of $\MI$ to $M_{\rm BH}$; this is captured by our $\sigma$ parameter. Even
if we knew global physical parameters for the core collapse process (e.g.,
reaction rates) with certainty, a given $\MI$ will always have a range of
possible associated remnant masses due to factors such as the unknown
metallicity at formation (see Sec.~\ref{sec:evolvingmassmodel}). 

In Fig.~\ref{fig:imf_mass_spectrum}, we show how the distribution of $M_{\rm
BH}$ is derived from an initial distribution of progenitor masses $\MI$
according to our model with some fiducial values. Each BH in the 1G population
is assumed to come from a progenitor from the $dN/d\MI$ distribution (bottom
panel), which is mapped to a remnant BH mass through the $\MI-M_{\rm BH}$
relationship (upper right panel). Finally, the resulting BH mass distribution
$dN/dM_{\rm BH}$ is obtained by integrating this distribution in the upper right
panel across $\MI$, weighted by $dN/d\MI$ (Eq.~\ref{dNdM1g}). This differs from
the procedure in \cite{Baxter21}, as we directly infer the $dN/d\MI$ and
$p(M_{\rm BH} \mid \MI)$ distributions, which uniquely specify $dN/dM_{1G}$
rather than inferring a phenomenological representation of a resulting
$dN/dM_{1G}$ distribution.

\hfill
\subsection{Full Mass Distribution}\label{sec:nonevolvingmassmodel}

The model outlined in the previous section is only directly applicable to 1G
mergers, in which $M_{\rm BH}$ is the remnant mass from the core collapse of one
of the sources from $dN/d\MI$. Realistically, a catalog of observed
gravitational wave sources could also contain higher-generation merger
events---namely, events that involve BHs who are themselves remnants of previous
BH mergers, and therefore not of stellar origin---although it is commonly
assumed that these systems will only subdominantly contribute to the inferred
mass distribution \citep{Miller02, Rodriguez19, Kimball21, Gerosa21}. As the
component masses of these events will be approximately (slightly less than) the
sum of the masses of the BHs from its \textit{previous} mergers, the masses in
this population can exceed $M_{\rm BH, max}$ and will not follow the same
distribution as the 1G BHs.

\begin{figure*}
    \centering
    \includegraphics[width=0.4\textwidth]{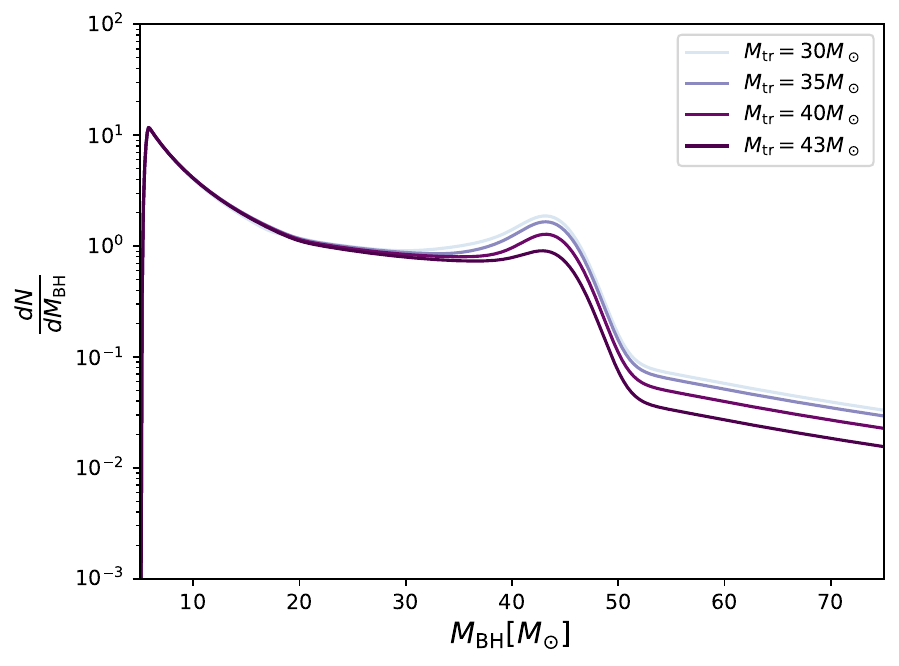}\qquad\qquad%
    \includegraphics[width=0.4\textwidth]{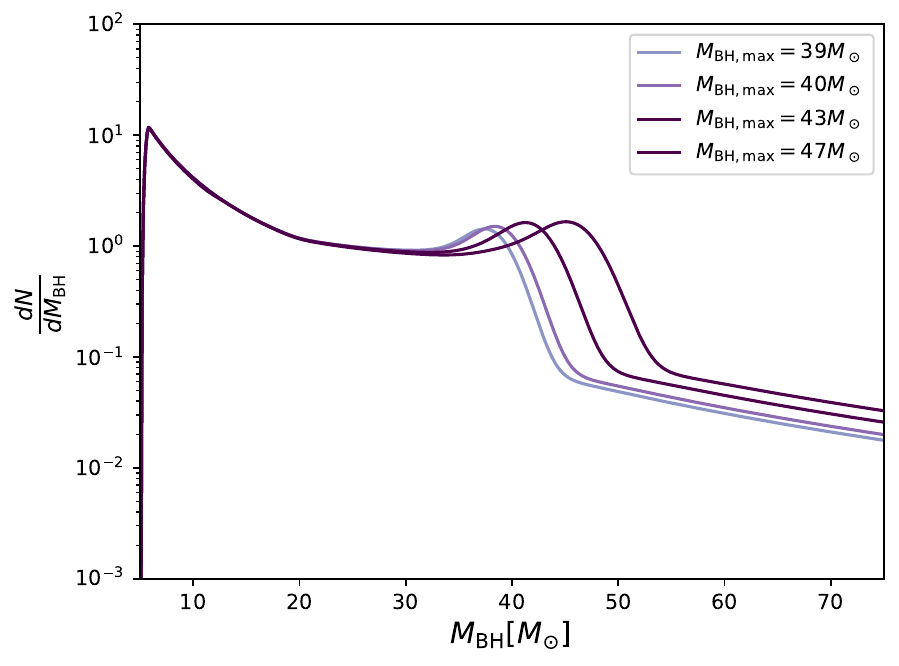}\\
    \includegraphics[width=0.4\textwidth]{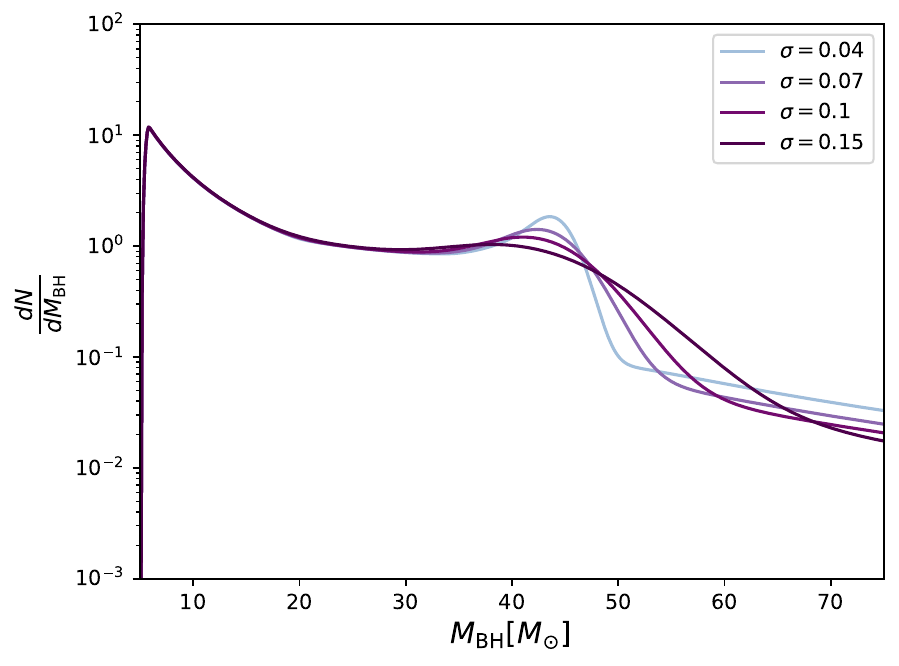}\qquad\qquad%
    \includegraphics[width=0.4\textwidth]{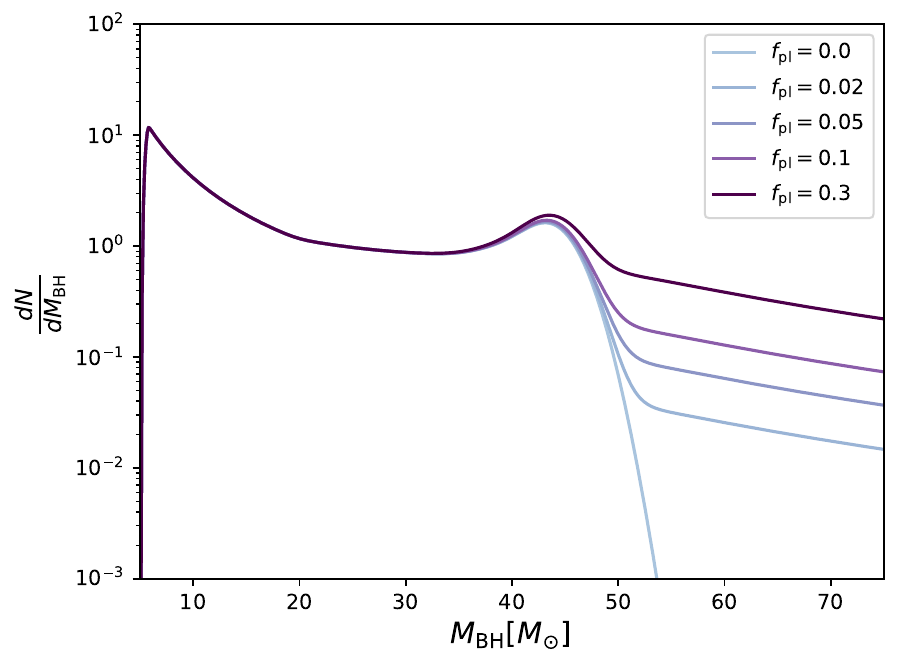}  
    \caption{BH mass spectrum following our model in Eq.~\eqref{fullmassdist} for different choices of (clockwise from top left) $\Mtr$, $M_{\rm BH,max}$, $f_{\rm pl}$, and $\sigma$. For the top left figure, we consider constant difference between $\Mtr$ and $M_{\rm BH,max}$. Unless being varied, we assume the following fiducial parameters: $a = 2, b = 1, c = 2.5, \Mtr = 35 M_\odot, M_{\rm BH,max} = 45 M_\odot, \sigma = 0.05$, and $f_{pl} = 0.04$.}
    \label{fig:mass_plots}
\end{figure*}

The details of the 2G distribution depend on unknown factors that make it
difficult to prescribe a specific functional form \citep{Rodriguez19, Kimball21,
Doctor20}. In order to capture these events in a relatively agnostic manner, we
enhance our model with a power law tail with a spectral index $c$ that smoothly
turns on just below $M_{\rm BH, max}$, and has a height $f_{pl}$ relative to
$dN/dM_{1G}$ at $M_{\rm BH,max}$ (see the bottom right panel in
Fig.~\ref{fig:mass_plots}). We express the full mass distribution as:

\begin{widetext}
\begin{equation}\label{dndm}
    \frac{dN}{dm} = \frac{dN}{dM_{1G}} + \delta(m \mid M_{\rm BH,max})\, f_{pl} \left.\frac{dN}{dM_{1G}}\right|_{M_{\rm BH,max}} \left(\frac{m}{M_{\rm BH,max}}\right)^{-c}  \, ,
\end{equation}
\end{widetext}

where $dN / dM_{1G}$ is given in Eq.~\eqref{dNdM1g} and $\delta(m)$ is an exponential tapering function that smoothly turns on to $M_{\rm BH,max}$;
the parameter $f_{pl}$ controls the relative height between the peak of the 2G power law and the 1G mass distribution at $M_{\rm BH,max}$.
By adopting this two-component model, we can prevent 2G sources from biasing the
inference of the parameters of the ${dN}/{dM_{1G}}$ distribution, which has a
sharp, log-normal falloff at masses $M_\mathrm{BH} > \MBHMax$. This assumes that
the 2G sources have a minimal contribution to the mass distribution below
${\sim} \MBHMax$, consistent with the conclusions from, e.g., \cite{Fishbach22}.

We model both component masses as coming from the same mass distribution ${dN}/{dM_{\rm BH}}$ and include a pairing function with power law slope $\beta$ to get the full mass distribution:
\begin{equation}\label{fullmassdist}
    \frac{dN}{dm_1 d m_2}(m_1,m_2) \propto \left(m_1 + m_2\right)^\beta \frac{dN}{dm_1} \frac{dN}{dm_2}  \, ,
\end{equation}
where each $dN/dm_{1/2}$ factor corresponds to a density as in Eq.~\eqref{dndm}.
The first factor in Eq.~\eqref{fullmassdist} constitutes the pairing function,
by which the component masses do not only inform the mass distribution
independently but also by how they pair together to form a total mass
\citep{Fishbach20, Farah23}. The parameter $\beta$ is the exponent on the total
mass, such that positive (negative) values for $\beta$ mean that masses pair up
to preferentially form systems of higher (lower) total mass.  We choose this
form of the pairing function, first suggested in \citet{Fishbach20}, to permit
the possibility of \emph{breaking} factorization symmetry, so that when $\beta
\neq 0$ the joint mass function is not the product of a function of $m_1$ and a
function of $m_2$; many of the models highlighted in \cite{gwtc3pops} are forced
to be symmetric in this sense.  

Fig.~\ref{fig:mass_plots} shows how the mass distribution $dN/dm$ changes as a
function of various population hyperparameters. The top left panel shows how
$\Mtr$ predictably controls onset of the transition to a peak in the mass
distribution by changing where the $\MI - M_{\rm BH}$ mapping becomes nonlinear;
additionally, as $\Mtr$ becomes closer to $\MBHMax$, the height of the peak
decreases as a smaller range of the IMF is contributing to the peak region. The
upper right panel shows how this peak moves to higher $M_{\rm BH}$ and gets
wider as $M_{\rm BH,max}$ increases for fixed $\Mtr$. The bottom left panel of
Fig.~\ref{fig:mass_plots} shows the effect of varying $\sigma$ on the resulting
mass distribution. As $\sigma$ increases, the remnant masses for a given core
mass scatter more broadly, smoothing the remnant mass distribution and softening
the peak, as well as making the cutoff above $\MBHMax$ weaker. Together, these
physical parameters govern the location, strength, and width of the peak in the
BH mass distribution, as well as the strength of its cutoff. The final panel of
Fig.~\ref{fig:mass_plots} demonstrates the increasing contribution of the
high-mass power law tail when raising $f_{pl}$.

\subsection{Redshift Model}\label{sec:redshift}

Studies of cosmic star formation history with astronomical surveys show that the star formation rate increases to a redshift of $z \approx 2$, then smoothly decays at high redshifts, which is well-modeled by a smoothly broken power law \citep{Madau14, Vangioni15, Ghirlanda16}. When convolved with a reasonable delay-time distribution, this also gives rise to a smoothly broken power law for the merger rate $R(z)$ \citep{Fishbach18}, i.e., the number of mergers per comoving volume ($V_c$) per time interval in the source frame ($t_s$).
Accordingly, we assume a redshift distribution such that
\begin{equation}\label{Rz}
    \frac{dN}{dV_c dt_s}(z) \equiv R(z) \propto \frac{\left(1 + z\right)^\lambda}{1 + \left(\frac{1+z}{1+z_{\rm peak}}\right)^\kappa} \, ,
\end{equation}
where, $\lambda$ controls the low-redshift merger rate, estimated to be $\lambda \approx 3$ in current LIGO/Virgo studies \citep{gwtc3pops}; meanwhile, the parameter $z_{\rm peak}$ controls the redshift at which the merger rate peaks and the slope becomes negative.

Both the peak redshift and the high-redshift merger rate are expected to be directly informed from detections beyond the horizon of current ground-based detectors or at redshifts where detections are scarce \citep{Callister23, Vitale19, gwtc3pops}. However, combining upper-limits from stochastic gravitational wave searches with population inference studies can place limits on these parameters with current detections \citep{Callister20}. Future observations with 3G detectors will allow us to significantly constrain the merger rate history across cosmic time using direct detections of BBH mergers at nearly all relevant redshifts \citep{Ng21, Mancarella23}.

The merger rate as a function of redshift in the detector frame is expressed simply in terms of mergers per redshift $z$ per detector-frame time $t_{\rm det}$:
\begin{equation}
    \frac{dN}{dzdt_{\rm det}}(z) = \frac{dN}{dV_c\, dt_s} \frac{dV_c}{dz} \frac{dt_s}{dt_{\rm det}} = R(z) \frac{dV_c}{dz}\frac{1}{1+z} \, ,
\end{equation}
where $R(z)$ is as in Eq.~\eqref{Rz}, and $dV_c/dz$ is the differential comoving volume per redshift bin as determined by cosmology.

\subsection{Allowing the Mass Spectrum to Evolve with Redshift}\label{sec:evolvingmassmodel}

Studies of stellar evolution predict that stars formed in lower-metallicity
environments can reach higher remnant BH masses before hitting the PISN cutoff.
This is generally attributed to the ability for metal-rich stellar winds to
carry off significant mass, resulting in lower remnant BH masses after
undergoing pulsations \citep{VanSon22, Marchant19, Farmer19}. While we do not
get direct information about the progenitor metallicities of gravitational-wave
sources from the observed data, we can use known correlations between
metallicity and observables in gravitational-wave data to look for this
evolution. Redshift and metallicity are anticorrelated: stars formed earlier in
the universe (i.e., at higher redshift) are metal-poor when compared to those
formed more recently (at lower redshift), due to the need for the existence of
pre-existing stars to deposit metals into the interstellar medium in order to
birth further generations of stars with higher metallicities \citep{Maiolino08,
Belczynski16}.

Several previous studies have used this trend as motivation to search for redshift-dependence in the observed BBH mass distribution.\footnote{\cite{VanSon22} proposes that differing delay time distributions between the high and low mass portions of the mass distribution may also result in an evolving mass distribution. Unlike evolution due to birth metalicity, this trend would not be tracked by the evolution in our model.} These studies have typically adopted phenomenological approaches to modeling this effect, directly encoding redshift dependence in the location of features in the BBH mass distribution.
Such features include the location of the Gaussian peak and the truncation point of the mass distribution, allowing these features to vary, for example, linearly with redshift or with some function of expected metallicity at a particular redshift \citep{Safarzadeh19, Fishbach18}.

We can leverage the physical framework we introduced in Sec.~\ref{sec:nonevolvingmassmodel} to model the redshift dependence in the mapping from $\MI$ to $M_{\rm BH}$. This allows us to treat the redshift evolution in the observed BBH mass distribution as a derived byproduct from an astrophysical process expected to evolve with redshift, rather than encoding the redshift dependence in the BBH mass distribution directly. 

We express this evolution in term of a linear expansion for the location of the $\Mtr$ turnover in the mass distribution:
\begin{equation}\label{mpisndot}
    \Mtr(z) = \Mtr(z = 0) + \Mtrdot \left( 1 - \frac{1}{1+z}\right),
\end{equation}
where $\Mtr$ and $\Mtrdot$ are free parameters which we can interpret as the
transition location at $z = 0$ and the change in this location over a Hubble
time, respectively. In order to maintain the constraint that $\MBHMax > \Mtr$, we apply an equivalent adjustment to $\MBHMax$ to maintain 
\begin{equation}
    \MBHMax(z) - \Mtr(z) = \mathrm{const}
\end{equation}
at all redshifts.  This is an indirect model of the evolution of this feature
from high metallicity (late universe, $z = 0$) to low-metallicity (early
universe, $z = \infty$) environments. One could alternatively construct a more
explicit model for $\Mtr$ and $\MBHMax$ as a function of metallicity and then
metallicity as a function of redshift. 

\begin{figure}
    \centering
    \includegraphics[width=0.48\textwidth]{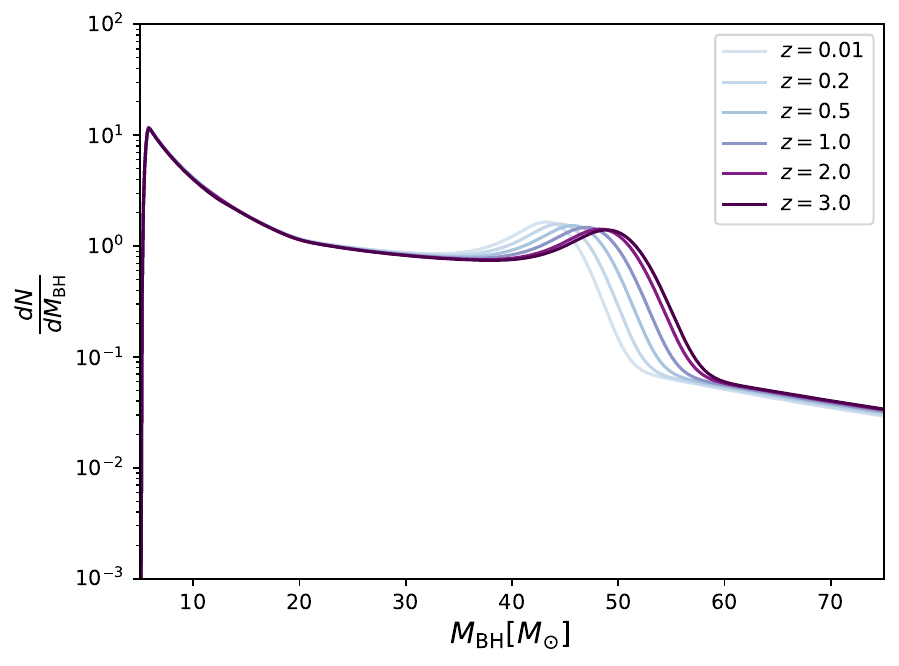}
    \caption{Model for a redshift-dependent mass distribution evaluated at selected redshifts assuming $\dot{M}_{\rm PISN} = 5\, M_\odot$ in Eq.~\eqref{mpisndot}. All other parameters are the same as the fiducial set in Fig.~\ref{fig:mass_plots}.}
    \label{fig:mass_evolvempisn}
\end{figure}

Fig.~\ref{fig:mass_evolvempisn} shows how our model for $dN/dM_{1G}$ appears for different values of redshift for two choices of $\Mtrdot$. To be consistent with predictions from stellar models, as described above, we expect a positive value for $\Mtrdot$ such that the turnover to PPISN occurs at higher masses at higher redshifts; equivalently, this means we expect the bump in the mass distribution moves toward higher mass at higher redshifts.

\section{Results}\label{sec:results}

\hfill
\subsection{Non-Evolving Mass Distribution}\label{sec:gwtc3nonevolvingresults}

Adopting the non-evolving mass and redshift models introduced in the previous section, we infer the corresponding hyperparameters using the hierarchical Bayesian inference approach described in Sec.~\ref{sec:hierarchicalinference} and the priors in Table \ref{tab:priors}. 

\begin{table*}
    \begin{center}
    \begin{tabular}{||l l l||} 
     \hline
     \textbf{Parameter} & \textbf{Prior} & \textbf{Description}\\ [0.5ex] 
     \hline\hline
     $a$ & $\mathcal{N}(2.35, 2) [-1.65, 6.35]$ & Power law index of low-mass CO IMF \\
     \hline
     $b$ & $\mathcal{N}(1.9, 2) [-2.1, 5.9]$ & Power law index of high-mass CO IMF \\
     \hline
     $c$ & $\mathcal{N}(4, 2) [0, 8]$ & Power law index of 2G high-mass tail\\
     \hline
     $\Mtr$ & $\mathcal{N}(35, 5) [20, 50]$ & Linear-to-quadratic transition mass $[M_\odot]$\\
     \hline
     $M_{\rm BH,max} - \Mtr$ & $\mathcal{N}(3, 2) [0.5, 7]$ &  Maximum remnant mass produced by the 1G channel, relative to $\Mtr$ $[M_\odot]$\\
     \hline
     $\sigma$ & $\mathcal{N}(0.1, 0.1) [0.05, ]$ & Width of lognormal distribution for $\MI$ to $M_{\rm BH}$ mapping\\
     \hline
     $\beta$ & $\mathcal{N}(0,2)$ & Exponent on total mass pairing function\\
     \hline
     $\log(f_{pl})$ & $\mathcal{U}[0.01, 0.5]$ & Log of relative height between the start of the 2G powerlaw and end of ${dN}/{dM_{1G}}$ \\
     \hline
     $\lambda$ & $\mathcal{N}(2.7, 2) [-1.3, 6.7]$ & Exponent controlling $R(z)$ at low redshift\\
     \hline
     $\kappa - \lambda$ & $\mathcal{N}(2.9,2)[1, 6.9]$ & Exponent controlling $R(z)$ at high redshift \\
     \hline
     $z_{\rm peak}$ & $\mathcal{N}(1.9, 1)[0, 3.9]$ & Redshift at peak $R(z)$\\
    \hline
     $\Mtrdot$ & $\mathcal{U}[-2, 8]$ & Difference in $\Mtr(z)$ over Hubble time $[M_\odot]$\\[1ex]  
     \hline
    \end{tabular}
    \caption{Priors used in this work. $\mathcal{U}$ is a uniform distribution and $\mathcal{N}$ is a Gaussian distribution with mean and standard deviation specified in the parentheses. Numbers in square brackets are upper and lower bounds of the prior.}
    \end{center}
    \label{tab:priors}
\end{table*}    

\begin{figure}
    \centering
    \includegraphics[scale=0.53]{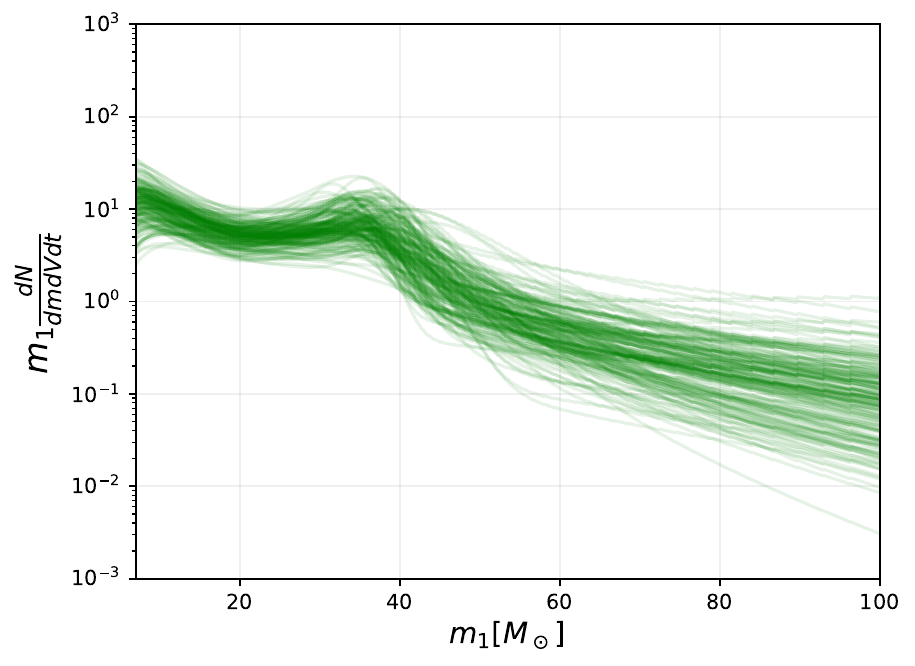}
    \caption{Draws from the non-evolving mass distribution posterior, evaluated at $z = 0$.}
    \label{fig:gwtc3traces}
\end{figure}

\begin{figure}
    \centering
    \includegraphics[scale=0.45]{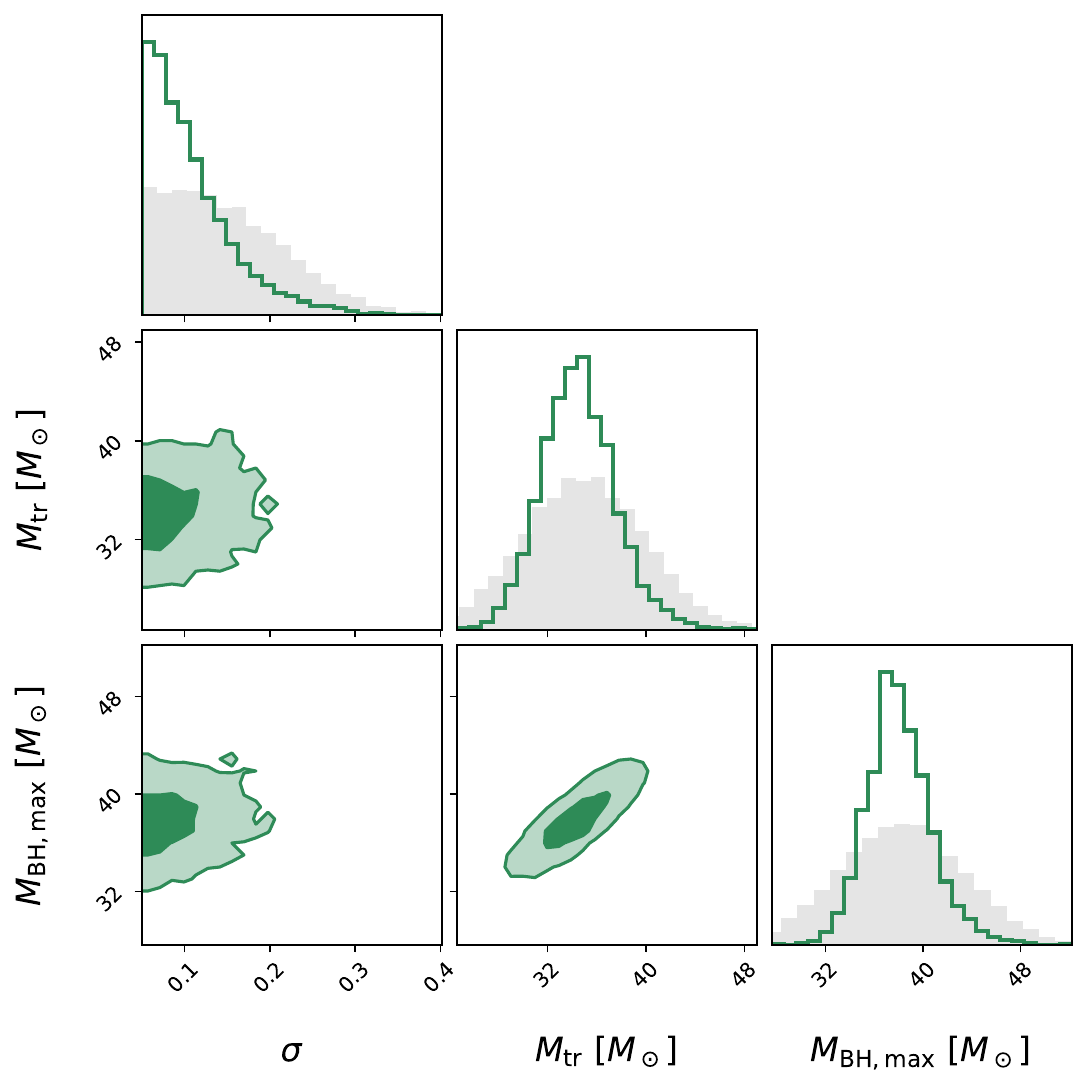}
    \caption{Posterior for selected mass distribution parameters using the model outlined in Sec.~\ref{sec:nonevolvingmassmodel}. Dark- and light-green shaded regions are the $1\sigma$ and $2\sigma$ contours, enclosing 39\% and 86\% of the probability respectively. Prior distribution is shaded grey for reference. We find that widening the prior does not meaningfully increase the posterior support for the PPISN feature at higher masses.}
    \label{fig:gwtc3sigmapisncorner}
\end{figure}

We plot draws from the mass distribution posterior in
Fig.~\ref{fig:gwtc3traces}, which shows the inferred decaying power law shape of
the mass distribution and the feature at ${\sim} 35 M_\odot$. Turning
attention to the parameters that most directly control the location and strength
of the peak in the mass distribution, Fig.~\ref{fig:gwtc3sigmapisncorner} shows
the posterior distributions for $\sigma$, $\Mtr$, and $M_{\rm BH, max}$. The
recovered distribution for $\sigma$ shows notable preference for low values,
converging toward the lower bound of the prior ($\sigma$ = 0.05). This means
that the data are consistent with little to no scatter around the $\MI-M_{\rm
BH}$ mapping, while ruling out high values of $\sigma$ that would over-smoothen
the peak in the mass distribution (cf.~Fig.~\ref{fig:mass_plots}, bottom panel).
By the same token, the strong support for low values of $\sigma$ indicates that
the data do allow for a relatively sharp cutoff in the peak; this is such that a
suppression of the high end of the peak need not be compensated by a higher rate
in the start of the 2G tail.

With the data supporting a peak at ${\sim} 35 M_\odot$, we measure a correlation between $\Mtr$ and $M_{\rm BH,max}$, driven by the constraint that $M_{\rm BH,max} > \Mtr$.
We note that the posteriors for $M_{\rm BH,max}$ and $\Mtr$ are different from the prior, indicating that the data are informing both the location and the width of the bump.

Taking the model at face value, we infer the location of $M_{\rm BH,max}$ to be
much lower than where stellar nucleosynthesis simulations generally predict the
upper mass gap due to the PISN process to begin; a similar conclusion was
reached in \citet{Farmer20}. For example, \cite{Farmer19, Farmer20} finds the
lower edge of the PISN mass gap to range between ${\sim} 45 {-} 50 \, M_\odot$,
when varying the CO reaction rate within its 1$\sigma$ uncertainty (with
standard deviation $\sigma_{C12}$) with respect to the distribution of reaction
rates given in \texttt{STARLIB} \citep{Sallaska13}. Using their fit to the start
of the mass gap as a function of $\sigma_{C12}$, and extrapolating down to our
inferred values of $M_{\rm BH, max}$, we infer $\sigma_{C12} =
4.8^{+3.1}_{-2.8}$ at 90\% credible levels.\footnote{For reaction rates this
high, the fraction of carbon in the core is too low ($X_{\rm C} \ll 10^{-3}$) to
be considered a CO core. To be consistent with the assumptions stated in
\cite{Farmer19}, $\MI$ can instead be interpreted as the mass within the
convective zone during helium burning.}  In other words, to match the location
of our observed peak would require a $\sim 5 \sigma$ adjustment in the C12
reaction rate relative to its current nuclear-physics best-estimate and
uncertainty.  Further discussion of this point can be found in Section
\ref{sec:discussion:ppisn}.

Although the simulation coverage is sparse at these masses and therefore these constraints are largely extrapolation-driven, the anomalous value inferred for this parameter casts doubt on this PPISN model as a an explanation for the ${\sim}35\,M_\odot$ peak in question. See Sec.~\ref{sec:discussion} for further discussion.

\begin{figure*}
    \centering
    \includegraphics[scale=0.4]{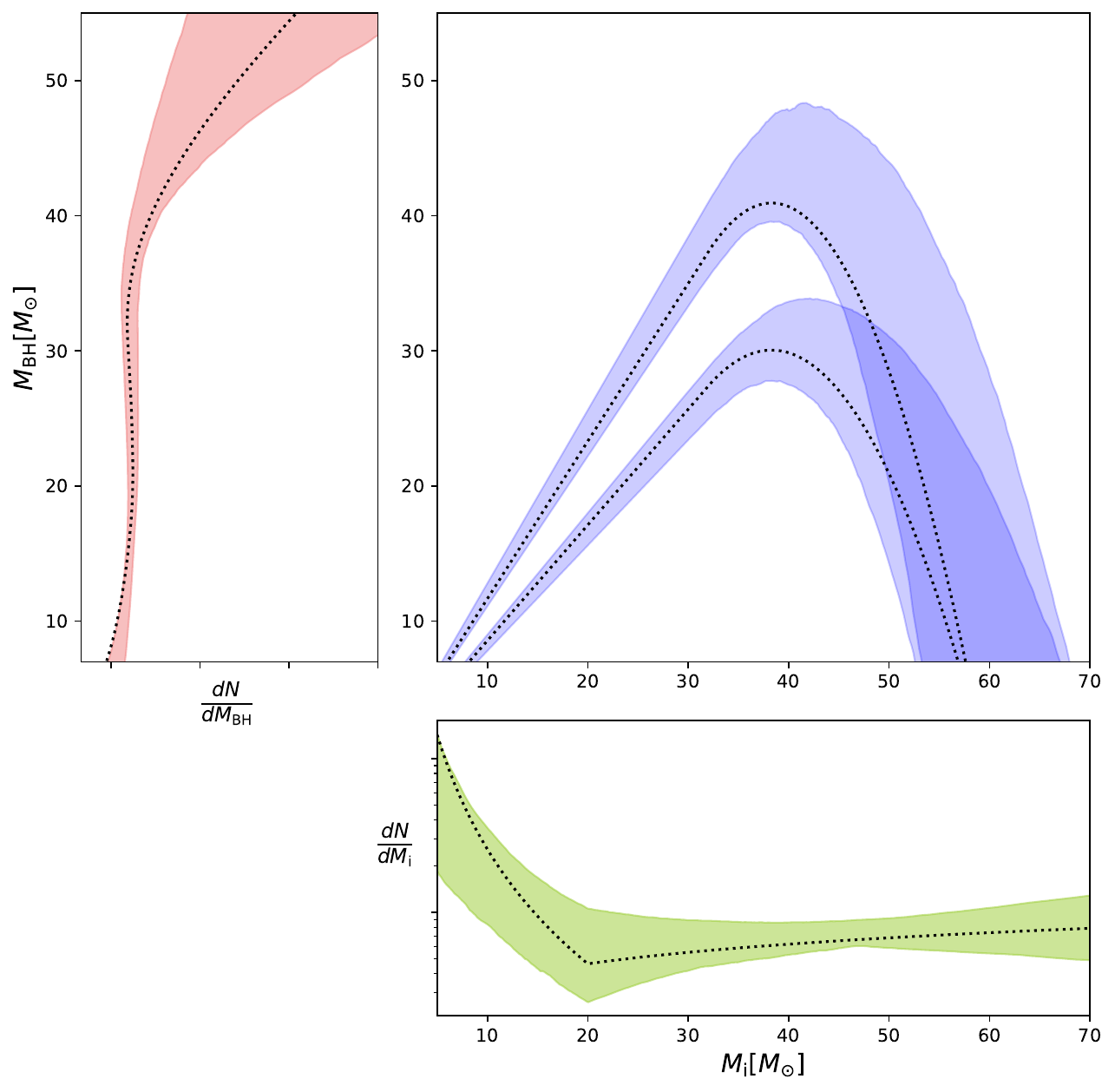}
    \caption{Representation of the $\MI - M_{\rm BH}$ relationship of the non-evolving $\frac{dN}{dM_{1G}}$ model in Sec.~\ref{sec:nonevolvingmassmodel}. (Bottom panel) Inferred distribution for the initial mass function of $\MI$ in merging binaries. (Top right panel) 1-$\sigma$ credible region of the 95th percentile (higher blue shaded region) and 5th percentile (lower blue shaded region) for the inferred $\MI - M_{\rm BH}$ mapping. Dotted line is a single representative draw from the posterior.}
    \label{fig:gwtc31gmassspec}
\end{figure*}

In Fig.~\ref{fig:gwtc31gmassspec} we plot the inferred inferred $\MI - M_{\rm
BH}$ mapping, the $\MI$ IMF, and 1G $M_{\rm BH}$ mass distribution. We find that
the $\MI$ IMF steeply decreases before the break at $\MI = 20 M_\odot$ and then
becomes shallower or flattens out; observations of massive stars in star forming
regions suggest that the high-mass IMF could be shallower than at lower masses
\citep{Schneider18}. The $\MI - M_{\rm BH}$ mapping is somewhat uncertain, but
the turnover reliably creates a peak at $\sim 35 M_\odot$. 

\subsection{Evolving Mass Distribution}\label{sec:GWTC3evolvingresults}

\begin{figure}
    \centering
    \includegraphics[scale=0.35]{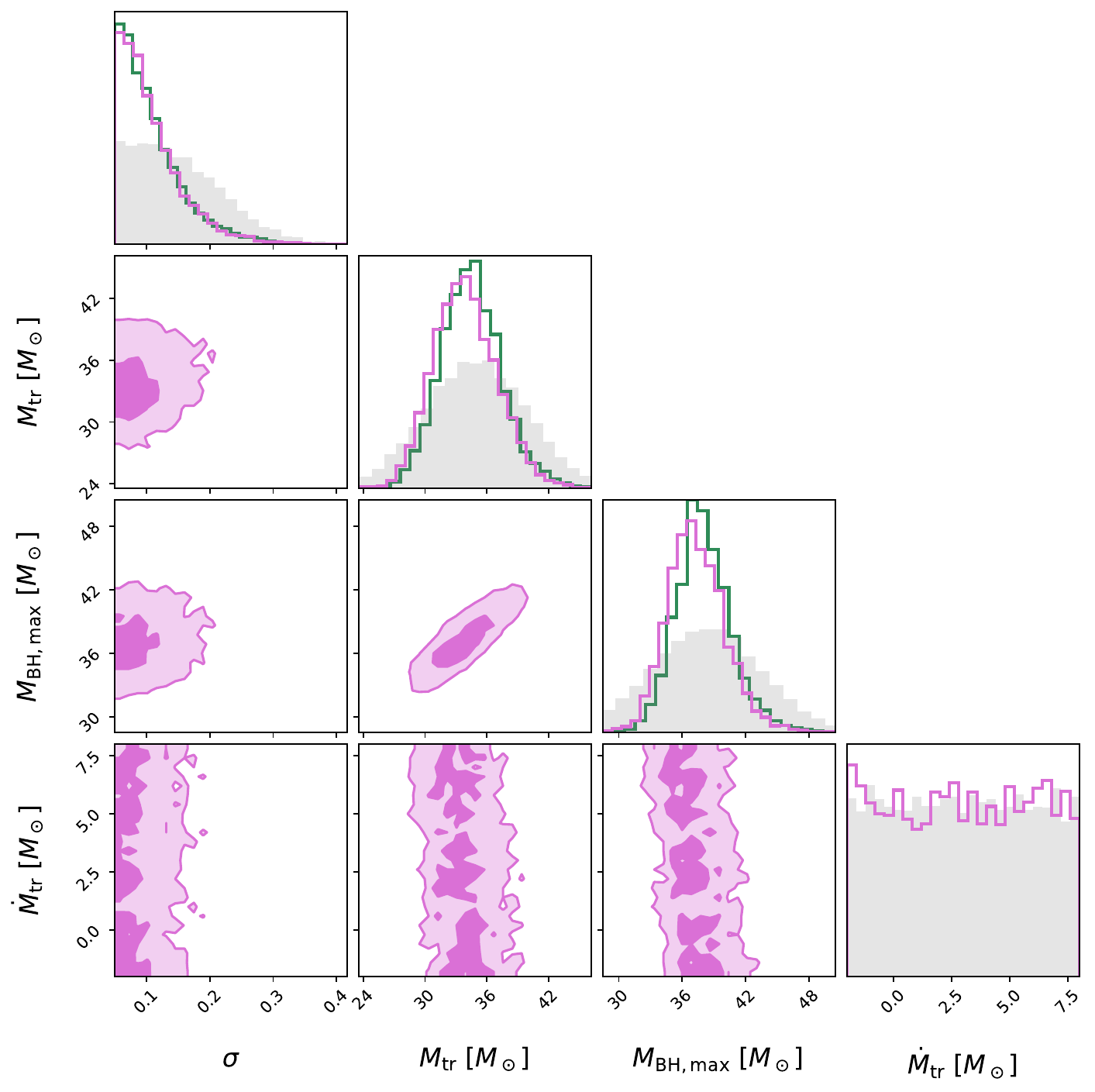}
    \caption{Posterior for selected mass distribution population parameters from the evolving mass distribution model in Sec.~\ref{sec:evolvingmassmodel} (magenta). One-dimensional posteriors from the non-evolving mass model overplotted in green for reference.}
    \label{fig:evolvingmassparameters}
\end{figure}

By adopting the more general model from Sec.~\ref{sec:evolvingmassmodel} we can relax some of the assumptions made in the previous sections and now infer the mass distribution in the presence of an $\Mtr$ that evolves with redshift. In Fig.~\ref{fig:evolvingmassparameters}, we present the posterior probability density on several mass distribution parameters from this model. Most of the events in the O3 catalog lie at relatively low redshift, and therefore do not provide good coverage across redshift scales to inform $\Mtrdot$ meaningfully. Due to these poor constraints on $\Mtrdot$, we find that the inferred distributions for $\Mtr$ and $M_{\rm BH,max}$ are consistent with those obtained when using the non-evolving model.
In other words, the feature at $\Mtr$ is being informed by structure in the data that does not appear to need to vary with redshift. The resulting mass distribution is consistent with that from the non-evolving model (plotted below in Fig.~\ref{fig:gwtc3compare}).

\begin{figure}
    \centering
    \includegraphics[scale=0.35]{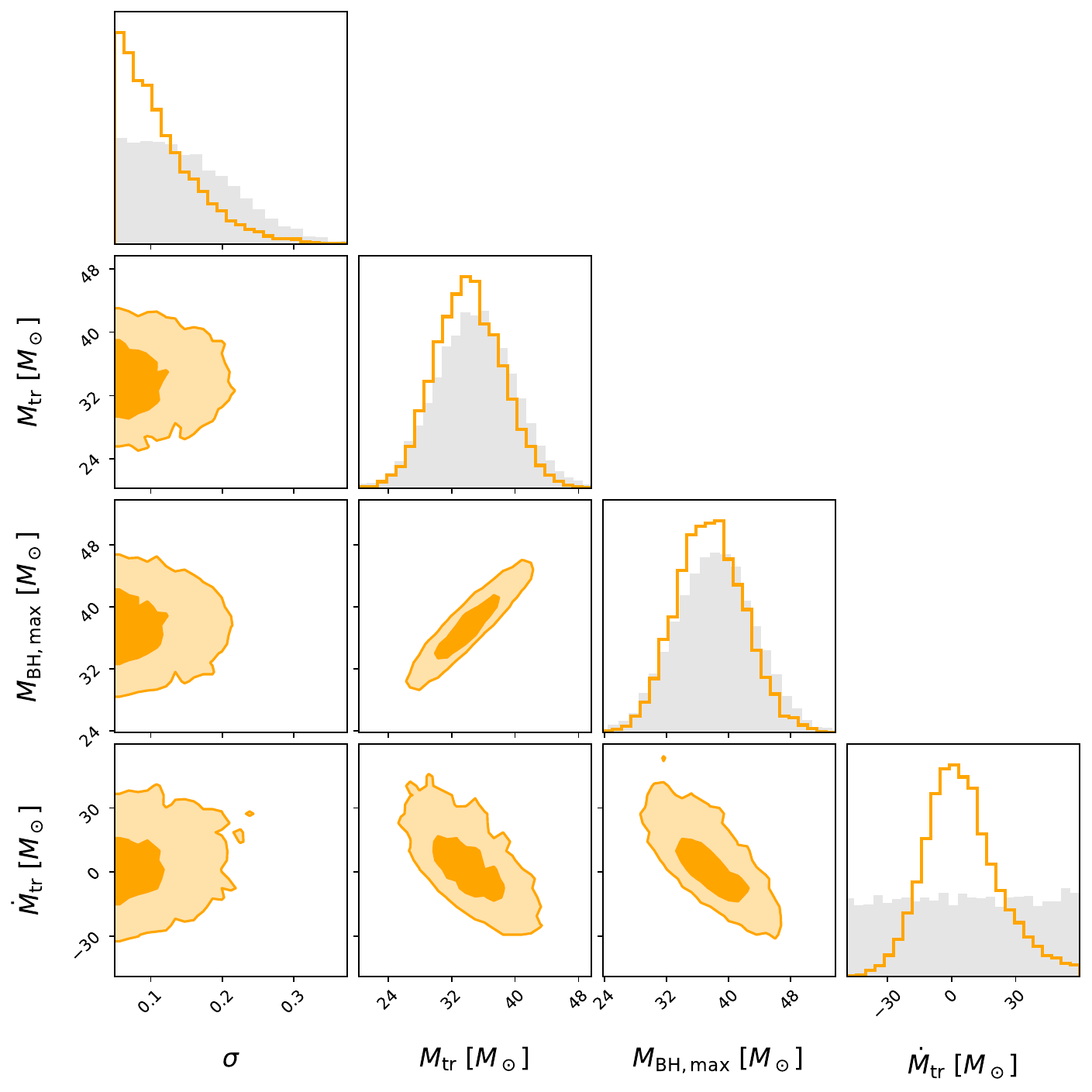}
    \caption{Inferred mass distribution parameters using the evolving mass model in Sec.~\ref{sec:evolvingmassmodel}, but adopting a wider prior on $\Mtrdot$. Prior distribution is shaded grey for reference.}
    \label{fig:mppisndotwiderprior}
\end{figure}

When we extend the prior on $\Mtrdot$ considerably, we find that we only rule
out redshift evolution of the intrinsic mass function for very large values of
$\Mtrdot$. In Fig.~\ref{fig:mppisndotwiderprior}, we show the posterior
distribution for relevant parameters: constraints on $\Mtrdot$ are broad,
encapsulating a 90\%-credible region from $\Mtrdot$ from $-20\, M_\odot$ to
$36\, M_\odot$.  This range is much broader than would be expected from the
metallicity dependence of the onset of pair instability pulsations
\citep{Farmer19}.  While this demonstrates that data rule out extreme values of
$\Mtrdot$, we cannot currently place constraints within a narrower, more
physically-relevant prior range; future observations may change this. The strong
anticorrelation between $\Mtrdot$ and $\Mtr$ likely indicates that we are
observing the peak from sources in a small range of redshifts, and we therefore
cannot constrain both free parameters in Eq.~\eqref{mpisndot}. In order to break
this degeneracy and get constraints on $\Mtrdot$, we would need additional
observations across redshifts. For the remainder of this section, we present
results using the narrower prior for $\Mtrdot$ (in Table \ref{tab:priors}).

\begin{figure}
    \centering
    \includegraphics[scale=0.45]{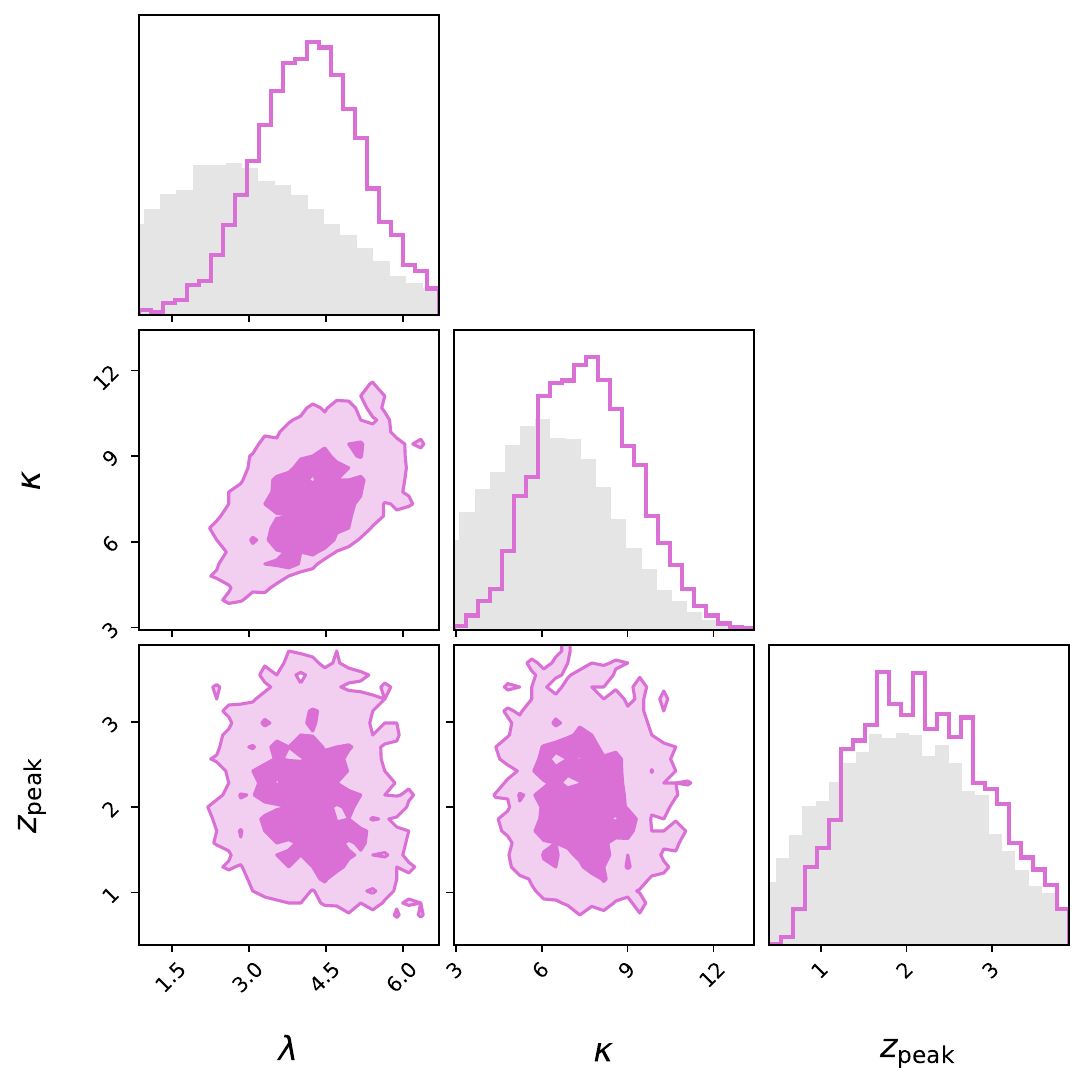}
    \caption{Redshift distribution parameters inferred with the evolving mass distribution model in Sec.~\ref{sec:evolvingmassmodel}. Prior distribution is shaded grey for reference.}
    \label{fig:evolvingmassredshiftparameters}
\end{figure}

Also of note is the similarity between the distribution of $\sigma$ obtained with this model and that obtained with the non-evolving mass model (see the comparison in Fig.~\ref{fig:evolvingmassparameters}). We discuss the implications of this in Sec.~\ref{sec:discussion}.

We present the distribution for redshift parameters, inferred jointly with the evolving mass distribution, in Fig.~\ref{fig:evolvingmassredshiftparameters}. 
The parameter best constrained is $\lambda$, which controls the evolution of the low-redshift merger rate. We infer $\lambda = 4.3^{+1.6}_{-1.6}$, preferring a merger rate that evolves \textit{steeper} than the low-redshift star formation rate ($\lambda \sim 2.7$). However, the evolution of the merger rate is still consistent with that implied by the star formation rate along with a short delay-time distribution. Narrower constraints on this parameter may reveal information on different formation channels contributing to the observed catalog of BBHs.

Additionally, the posterior distribution for $z_{\rm peak}$ is shifted slightly to the right of the prior, meaning that we are able to begin to place very conservative lower limits on $z_{\rm peak}$ due to the lack of a visible start of a turnover in the inferred $R(z)$ distribution. The lack of support at the tails towards higher $z_{\rm peak}$ is not due to information gained from the data, but rather from the prior (see Table \ref{tab:priors}). These constraints appear despite having only very little high-redshift information and are therefore very weak limits. 

Given the similarities between the inferred distributions with and without redshift evolution in the $\MI$ to $M_{\rm BH,max}$ map, we do not currently obtain improved constraints on physical parameters of interest when modifying the model in this way. For example, we infer $\sigma_{C12} = 5.2^{+3.4}_{-2.9}$, which is similar to what we reported in Sec.~\ref{sec:gwtc3nonevolvingresults}.
Future detections at higher redshift may further inform these aspects of the model.

\section{Interpretation of Results}\label{sec:discussion}

By adopting the model introduced in this work, we can draw conclusions from the inferred underlying physics represented in our models and explore how the population model compares to those reported in other works.

\begin{figure}
    \centering
    \includegraphics[scale=0.55]{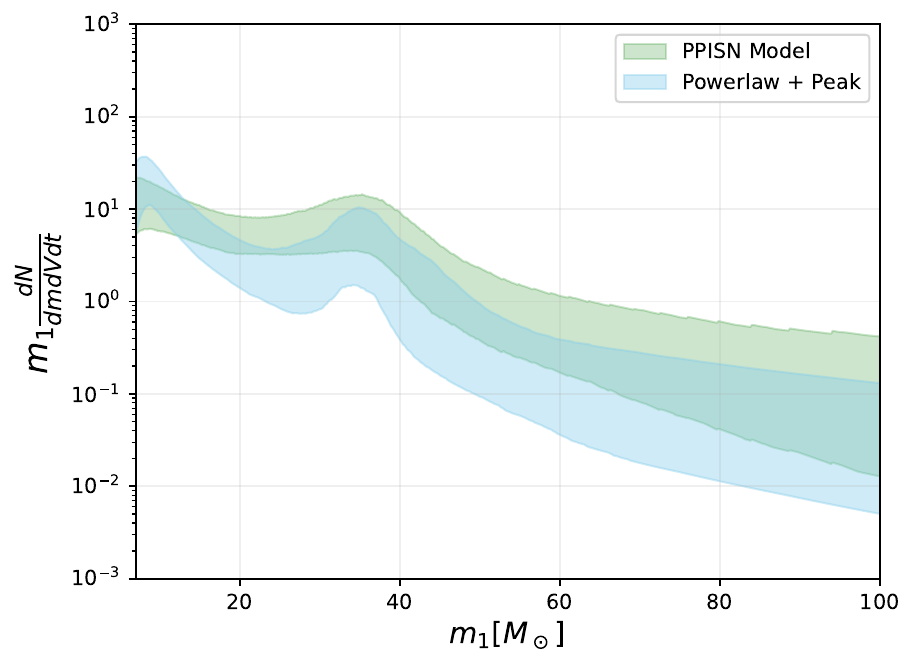}
    \includegraphics[scale=0.55]{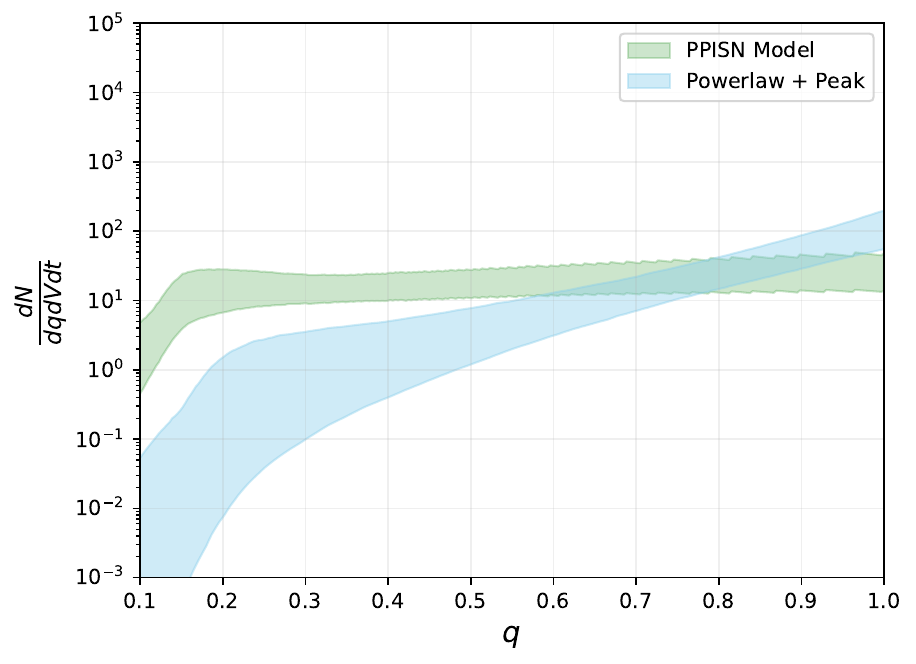}
    \caption{Mass spectrum derived from our model (green) with no evolution of the mass distribution with redshift compared to the \textsc{Powerlaw + Peak} mass spectrum informed by the same events (blue). We do not include redshift evolution of the mass distribution in this comparison as the models in \cite{gwtc3pops} do not include mass-redshift correlations. \textsc{Powerlaw + Peak} results were obtained using \texttt{GWPopulation} \citep{Talbot19}. (Top) Comparison of the primary mass distributions. (Bottom) Comparison of the mass ratio distributions.}
    \label{fig:gwtc3compare}
\end{figure}

\subsection{Global Shape of the Mass Distribution}

For comparison, we obtain results using the same set of O3 events, adopting the \textsc{Powerlaw + Peak} mass distribution model, a flat spin magnitude and tilt model, and a broken power law redshift distribution as implemented in \texttt{gwpopulation} \citep{Talbot19, TalbotThranePLP, Fishbach18}. Qualitatively, we infer a mass distribution (marignalized over $q$) consistent with the \textsc{Powerlaw + Peak} model, with major features such as the slope at higher BH mass as well as the bump location showing good agreement in Fig.~\ref{fig:gwtc3compare}. This indicates that this overdensity is a confident feature in the data whose location and prominence is not affected by systematic differences between these two models. This is reinforced by several other works, which find that models must include such a feature in order to faithfully capture the observed mass spectrum \citep{gwtc3pops, Farah23b, Edelman22,Callister23}.

A notable difference is in the merger rate relative to \textsc{Powerlaw + Peak}. While the 90\% credible regions overlap in Fig.~\ref{fig:gwtc3compare}, their relative heights show that our model tends to prefer a higher merger rate than what is predicted by \textsc{Powerlaw + Peak}, particularly at masses above ${\sim}15 M_\odot$.

In our model, $m_1$ and $m_2$ both directly inform the physical mass distribution model (along with the pairing function, see Eq.~\ref{fullmassdist}). This is in contrast to the \textsc{Powerlaw+Peak} model which has separate distributions for $p(m_1)$ and $p(q|m_1)$, such that $m_2$ does not directly inform $dN/dm_1$. As demonstrated in \cite{Farah23}, this makes the mass distribution feature a peak in the joint $m_1 - m_2$ space, rather than in the marginal $m_1$ distribution. Furthermore, in the marginal mass ratio distribution (bottom panel of Fig.~\ref{fig:gwtc3compare}), we see that our model prefers a much flatter distribution in $q$ than what is preferred by \textsc{Powerlaw + Peak}, which explicitly models $p(q|m_1)$ as a power law. This flat mass ratio distribution is consistent with what is found in \cite{Fishbach20} when adopting a pairing function that is a power law in $q$. Given that we infer $\beta$ with a preference for negative values, we find that BHs tend to pair up in binaries that favor lower total masses; this may cause a relative lack of support at higher masses in the mass distribution which models both ${dN}/{dm_1}$ and ${dN}/{dm_2}$. We also note that the inferred local rate $R(z = 0)$ is consistent between models. The \textsc{Powerlaw + Peak} model fits the underlying distribution (i.e., not including the bump) with a single power law, limiting the possible morphologies. We have checked that allowing the underlying power law in \textsc{Powerlaw + Peak} to include a break does not resolve the discrepancy.

The distribution for $p(\MI)$ we infer (see Fig.~\ref{fig:gwtc31gmassspec})
disagrees with what one may expect from an IMF resulting from the ZAMS mass IMF
assuming a linear relationship between ZAMS mass and $\MI$. While the
distribution is consistent with a decaying power law for low masses, the
distribution appears to flatten out above our break point of $20\, M_\odot$.
This trend is not strongly correlated with the $\Mtr$ and $M_{\rm BH,max}$ we
infer.

Comparing our results to those obtained in \cite{Baxter21}, we find strong tension with the maximum BH mass in the 1G channel (the start of the upper mass gap). Motivated by stellar evolution simulations to model the 1G BH mass distribution with a phenomenological approximation to the shape and location of a overdensity due to PPISN pileup, \cite{Baxter21} finds the PPISN feature and corresponding start of the upper mass gap to be at ${\sim} 46 M_\odot$, in very good agreement with predictions from typical values of the $\rm ^{12}C(\alpha, \gamma)^{16}O$ reaction rate. Notably, \cite{Baxter21} does not find the feature at ${\sim} 35\, M_\odot$ we find and is consistently found in the literature.

\subsection{Evolution of Mass Distribution with Redshift}

Our finding that $\Mtrdot$ is consistent with zero agrees with other studies that do not find strong preference for evolution of the BBH mass distribution with redshift. For example, \cite{Fishbach21} models the mass distribution as a broken power law where the mass at which the power law breaks is allowed to vary with redshift. While this is a very different model, it should qualitatively reproduce some of the features of our model, particularly at the ${\sim}35\,M_\odot$ feature (see Fig.~\ref{fig:mass_plots}). We therefore expect that if \cite{Fishbach21} had found strong preferences for an evolving mass distribution, we would confidently find $\Mtrdot > 0$. We also agree that the data are still consistent with a mass distribution that has some evolution with redshift, but again we do not have positive evidence that this is the preferred scenario.

\cite{Karathanasis23} also looks for evolution of the mass distribution with redshift. The authors allow the Gaussian bump in a \textsc{Powerlaw + Peak}-like model to vary with redshift, where the placement of this peak at a given redshift is determined by the delay time distribution and a jointly-inferred model for the evolution of (birth) metallicity with redshift. The value they find for the lower edge of the upper mass gap of ${\sim} 44 M\, _\odot$ is nominally in better agreement with the prediction from stellar physics models. However, this value is cited at low metallicity, and they also find there must be a very strong evolution of this mass scale with metallicity. Extrapolating their results to the local universe, they find that the upper mass gap at $z = 0$ starts at $ {\sim} 30\, M_\odot$, which is closer to the corresponding value we obtain for the start of the $\MI {-} M_{\rm BH}$ turnover. This result seems in tension with theoretical predictions given how small of an effect metallicity is expected to have on $\Mtr$. There are unexplained differences in our results, however, as such a strong evolution of $\Mtr$ with metallicity should mean that we would infer a positive $\Mtrdot$, assuming delays do not mix events from many different birth metallicities into similar merger times.

If metallicity evolution effects were causing some of the support for nonzero values of $\sigma$ obtained in the non-evolving mass model (i.e., from scatter in the $\MI - M_{\rm BH}$ relation), we would expect $\sigma$ to be constrained closer to zero with the evolving model, as some of that scatter would have been absorbed by the redshift evolution. Given that this is not the case, we conclude that either (1) birth metallicity effects fundamentally have a subdominant impact on the $\MI {-} M_{\rm BH}$ relationship compared to other physical parameters that vary between BBH systems, or that (2) the birth metallicities of the systems in our catalog are not strongly correlated with the redshifts at which they merge. The latter scenario could result from the delay time distribution between formation and merger having enough support in the long-delay tails such that we cannot yet discern a strong correlation between birth time and merger redshift for systems merging at redshifts of $z \lesssim 1$.

\subsection{Physical Interpretations: PPISN Process}
\label{sec:discussion:ppisn}
We can take advantage of the physical parameterization of our model to interpret the implied stellar physics, assuming $M_{\rm BH,max}$ corresponds to the maximum 1G BH mass as determined by the PPISN process at a given redshift or metallicity.

Under the PPISN model, the pileup in BH masses around ${\sim}35 M_\odot$ would result from the remnants of stars with zero-age main-sequence (ZAMS) masses between ${\sim} 60 M_\odot$ through ${\sim} 140 M_\odot$ \citep{Rahman22, Woosley21, Woosley17}, driven by nuclear processes in the core \citep{Fowler64, Rakavy67}
For stars with ZAMS masses above this range, similar processes 
completely disrupt the star, leaving behind no remnant. Since the PPISN 
process produces a small range of remnant BH masses from stars from a wide range of ZAMS masses, it is expected that the mass distribution will exhibit the bump due to this pileup (sometimes referred to as the ``PPISN graveyard") followed by a suppression of sources, known as the upper mass gap \citep{Woosley17, Woosley19, Woosley21, Farmer19}.

Simulations of stellar evolution \citep[e.g.,][]{Farmer19, Mehta22, Farag22} have explored the relationship between initial stellar mass (in particular the mass of the Carbon-Oxygen (CO) core, $\MI$) and the final BH mass ($M_{\rm BH}$) after core collapse. They have also quantified the dependence of the location of the lower edge of this mass gap and its associated mass range on other physical parameters such as nuclear reaction rates, metalicity, and details of neutrino physics. Previous studies have used this relationship to place constraints on the astrophysical properties of the pulsational pair instability process, assuming the most massive sources observed through LIGO are below the upper mass gap \citep{Farmer20, Mehta22, Farag22, Stevenson19}. \cite{Baxter21} instead infers the population of BHs coming from the first-generation (1G) subpopulation below the upper mass gap along with the subpopulation of higher-generation (2G+) BHs (i.e., BHs that are themselves the product of past mergers) whose masses can lie within the upper mass gap.

\begin{figure}
    \centering
    \includegraphics[scale=0.42]{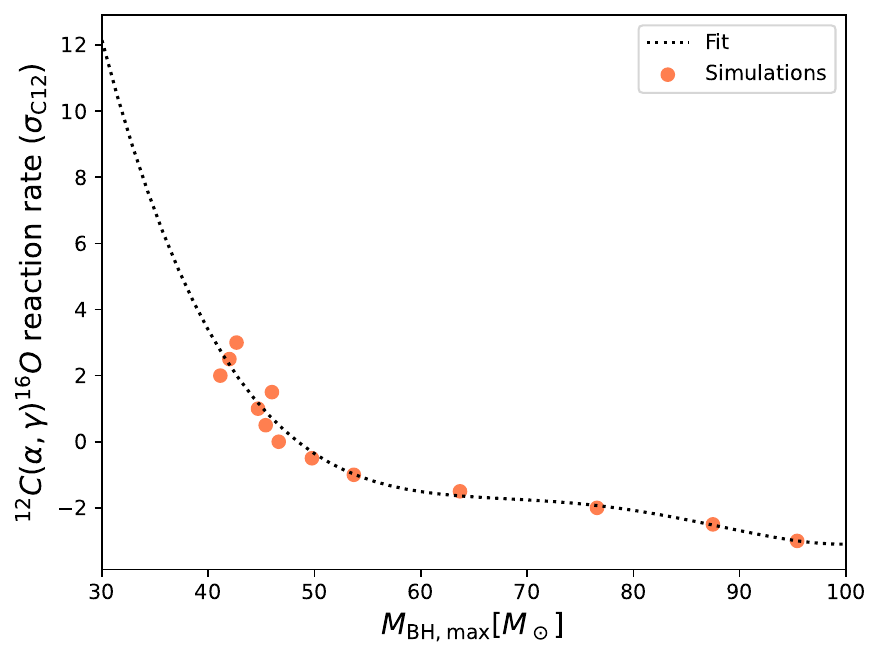}
    \includegraphics[scale=0.42]{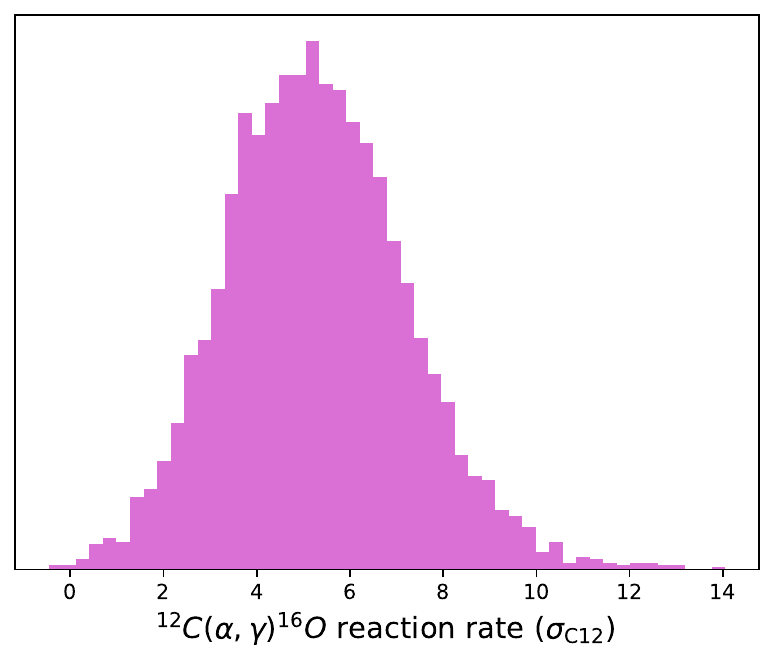}
    \includegraphics[scale=0.42]{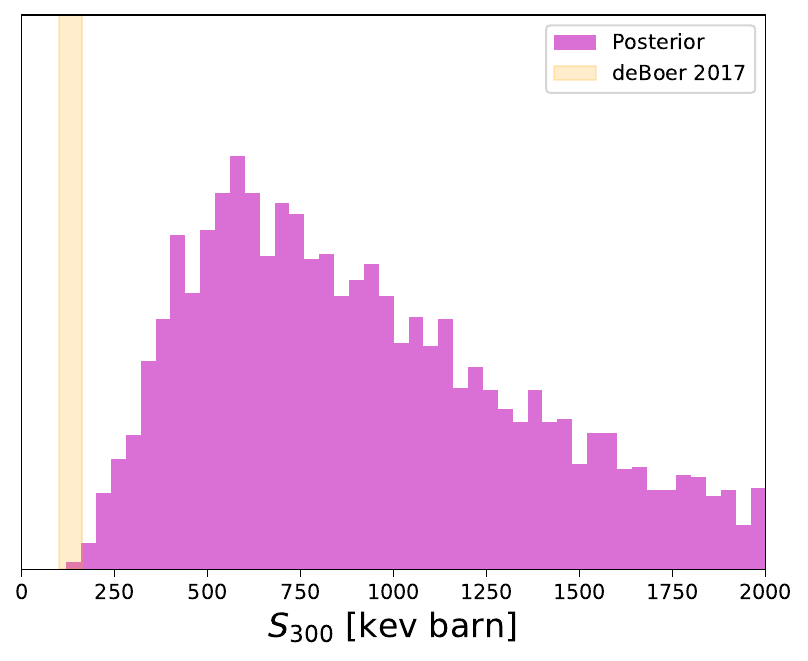}
    \caption{ (Top) Relationship between $\sigma_{C12}$ and lower edge of upper mass gap, reproduced from data release of \cite{Farmer20}. (Center) Posterior distribution of $\rm ^{12}C(\alpha, \gamma)^{16}O$ reaction rate, in terms of standard deviations away from the median reaction rate in \texttt{STARLIB} \citep{Sallaska13}; computed by evaluating the fit in the top panel for the samples of $M_{\rm BH,max}$ in the posterior in Fig~\ref{fig:evolvingmassparameters}. (Bottom) Inferred distribution of $S_{300}$, extrapolated from distribution of $\sigma_{C12}$ as calculated from the fit in the top panel. Constraints on $S_{300}$ from \cite{deboer17} plotted for comparison, showing tension with the values implied from our results.}
    \label{fig:bh2sig}
\end{figure}

Based on simulations \citep{Farmer20, Mehta22, Farag22}, the $\rm ^{12}C(\alpha,
\gamma)^{16}O$ reaction rate is likely to be the dominant physical factor
controlling $M_{\rm BH, max}$. Under this assumption, Fig.~\ref{fig:bh2sig}
shows the fit for $M_{\rm BH, max}$ as a function of $\sigma_{C12}$ (i.e., the
number of standard deviations from the median reaction rate in
\cite{Sallaska13}, in turn, adopted from  \cite{Kunz02}), reproduced from the
data release in \citet{Farmer20}. While their simulations only cover the range
$-3 < \sigma_{C12} < 3$, there is a clear trend that $\sigma_{C12}$ must rise
very steeply to reach a maximum BH mass below ${\sim} 45\, M_\odot$. We offer
details of this trend in Appendix~\ref{app:ppisndetails}.

After translating our inferred $M_{\rm BH,max}$ into  $\sigma_{C12}$ via the top of Fig.~\ref{fig:bh2sig}, we use the method in \cite{Farmer20} to estimate corresponding $S$-factors.\footnote{This ``astrophysical $S$-factor" is the part of the cross section given by the matrix element for the nuclear reaction itself, ignoring Coulomb repulsion
\citep[see][]{Kippenhahn94}.}
We arrive at a value of the $S$-factor at $300\, {\rm keV}$ of $S_{300} = 932^{+1929}_{-581}\, \mathrm{keV \cdot barn}$.

In Fig.~\ref{fig:bh2sig}, we compare this estimate of $S_{300}$ to the value predicted from nuclear experiments in \cite{deboer17}. That value is in strong tension with our estimate, which rules it out at $> 99.9\%$ credibility.%
\footnote{This is in contrast with \cite{Farag22} and \cite{Farmer20}, which infer $S_{300}$ to be consistent with \cite{deboer17}. However, those studies only consider the most massive observed BHs and assume they are 1G BBH mergers, without allowing for contamination from possible 2G mergers.}
Given this, we conclude that at least one of our assumptions is invalid.

One assumption is that the $\rm ^{12}C(\alpha, \gamma)^{16}O$ reaction rate is the physical parameter behind $M_{\rm BH, max}$. Multiple studies find that varying the $\rm ^{12}C(\alpha, \gamma)^{16}O$ reaction rate has a much stronger effect on the location of the start of the upper mass gap than other relevant reaction rates 
\citep{Farmer20, Farag22}. As these simulations only go up to $\sigma_{C12} = 3$, we cannot confirm the effect that varying other reactions rates when $\sigma_{C12}$ is high has on the location of the lower edge of the upper mass gap. While it is possible that one of these other reaction rates can be varied within their uncertainties to allow us to infer a lower $\sigma_{C12}$, it would have to change the location of the upper mass gap substantially to agree with the data \citep{Farmer19, Farag22, Woosley21}. 

Other assumptions inherent in present models of the PPISN process that could
affect the maximum BH mass include the treatment of convection
\citep{Renzo2020Convection} and the hydrodynamic treatment of mass ejection from
pair instability pulses \citep{Renzo2020Ejecta}.

Yet another, more fundamental, assumption in this interpretation is that the
turnover in the $\MI - M_{\rm BH}$ mapping can be associated with the
pair-instability process at all. 

In fact, our measurement in Fig.~\ref{fig:bh2sig} suggests that the observed bump in the mass distribution is not due to the PPISN turnover.\footnote{Mass-loss prescriptions and temporal resolutions of the simulations may introduce an unknown systematic bias \citep{Mehta22, Farag22, Farmer19}; however varying these settings has a subdominant impact on the location of the upper mass gap in simulations.}
Previous studies have suggested that associating the observed peak in the mass distribution with the PPISN pileup would be in tension with known stellar physics and observed supernovae rates \citep{Hendriks23, Woosley21}. Inferring the underlying $\MI - M_{\rm BH}$ mapping that gives rise to the observed BH mass distribution, this work also provides evidence of such a tension in terms of the underlying physics that would be necessary to generate a turnover in the $\MI - M_{\rm BH}$ map at the correct location.
Our model allows us to directly infer this tension from the GW data.

The cause of the peak in the observed mass distribution at ${\sim} 35\, M_\odot$
may therefore requires alternative explanations \citep{Hendriks23}. Recent
studies have proposed that this overdensity could be a signature from a
subpopulation of binaries which had undergone stable mass transfer
\citep{Briel23}, BBH systems in globular clusters \citep{Antonini23}, and stars
which have undergone significant wind-driven mass loss. Our model could be used
to describe any mechanism that generates a peak in the high-mass tail of the 1G
mass distribution via a transition to a nonlinear $\MI - M_{\rm BH}$
relationship.

\subsection{Model Limitations}

Our model has a few additional caveats. For example, our model does not attempt
to fit for the features beyond a power law that we know exist at lower masses
\citep{Edelman22, gwtc3pops, Farah23b}. We have confirmed that neglecting this
does not bias the inference in the higher-mass region that we care about here,
and this will be explored further in future work. We also ignore the effect of
spins in our population, but we demonstrate in Appendix~\ref{app:spins} that
this does not cause a notable bias in our results of interest. Given that
certain mass-spin correlations have been found in the BBH population
\citep{Callister21}, it may be insightful to use the spins to help distinguish
the 1G and 2G subpopulations (see, e.g., \cite{Fishbach17b, Gerosa17, Farr17}.

\section{Conclusions}\label{sec:conclusions}

Characterizing the population of BBH masses with direct phenomenological or nonparametric fits can provide insight into the shape of the mass distribution, but does not provide direct constraints on the underlying physics of BBH masses. With the method we propose here, we can infer the underlying physics by fitting the implied (derived) astrophysical BBH distribution to the observed data. We demonstrate the use of this method by evaluating the role of the PPISN process giving rise to the 1G BH mass distribution and its structure, including an excess (bump) in the mass distribution at the lower edge of the PPISN mass gap.
Fitting this model to the observed data, we find that the necessary physical parameters to explain the excess of BHs at ${\sim}35\, M_\odot$ are unrealistic from a nuclear physics perspective if we take this PPISN model at face value. We therefore conclude it is highly unlikely that the feature at ${\sim} 35\, M_\odot$ is associated with the PPISN process.

This framework motivates future investigations to better constrain the physics underlying astrophysical populations in general. Future work using additional observations and enhanced versions of our model may be able to constrain proposed astrophysical mechanisms underpinning the BBH mass, spin, and redshift distributions. This approach may offer fruitful applications such as calibration of ``spectral siren" features for cosmology \citep{Farr19b}, investigating other proposed interpretations of the bumps in the mass distribution, and understanding progenitor populations by relating back to population synthesis configurations \citep{Wong22, Andrews21, Zevin17}. 

\section*{Acknowledgements}
We thank Mathieu Renzo for helpful discussions about \texttt{MESA} simulations and the physics of stellar evolution and the PPISN process. We also acknowledge Alan Weinstein for useful comments and discussions, and Maya Fishbach for helpful comments on the manuscript.
The Flatiron Institute is a division of the Simons Foundation.
The authors are grateful for computational resources provided by the LIGO Lab and supported by National Science Foundation Grants PHY-0757058 and PHY-0823459. JG is supported by NSF award No. 2207758.
This research has made use of data or software obtained from the Gravitational Wave Open Science Center (gwosc.org), a service of the LIGO Scientific Collaboration, the Virgo Collaboration, and KAGRA. 

This material is based upon work supported by NSF's LIGO Laboratory which is a major facility fully funded by the National Science Foundation, as well as the Science and Technology Facilities Council (STFC) of the United Kingdom, the Max-Planck-Society (MPS), and the State of Niedersachsen/Germany for support of the construction of Advanced LIGO and construction and operation of the GEO600 detector. Additional support for Advanced LIGO was provided by the Australian Research Council. Virgo is funded, through the European Gravitational Observatory (EGO), by the French Centre National de Recherche Scientifique (CNRS), the Italian Istituto Nazionale di Fisica Nucleare (INFN) and the Dutch Nikhef, with contributions by institutions from Belgium, Germany, Greece, Hungary, Ireland, Japan, Monaco, Poland, Portugal, Spain. KAGRA is supported by Ministry of Education, Culture, Sports, Science and Technology (MEXT), Japan Society for the Promotion of Science (JSPS) in Japan; National Research Foundation (NRF) and Ministry of Science and ICT (MSIT) in Korea; Academia Sinica (AS) and National Science and Technology Council (NSTC) in Taiwan.

This manuscript carries LIGO Document Number \#P2300418.

\appendix
\section{Details of the Likelihood and Differential Rate Calculation}\label{app:derivation}

Setting $\theta$ to be the set of single-event parameters,
we can write the contribution from i$th$-event to the population likelihood as \citep{Mandel19}:

\begin{equation}\label{pereventlikelihood}
    p(d_i|\Lambda) = \frac{\int d\theta_i\, p(d_i|\theta_i) \, p(\theta_i|\Lambda)\, p_{\rm det}(\theta_i, d_i)}{\int\int dd_i\, d\theta_i\, p(d_i|\theta_i) \, p(\theta_i|\Lambda)\, p_{\rm det}(\theta_i, d_i)}
\end{equation}
Recalling that the probability density should be normalized over the arguments on the left side of the bar, the denominator is included to explicitely normalize the numerator in terms of the data from the i$th$ detection, and is commonly known as the ``selection effects" term. We write the detection probability as $p_{\rm det}(\theta_i, d_i)$ in order to include the general possiblity of thresholding detection in terms of the event parameters, which may be implemented when considering, for example, a simulated catalog. For our purposes, the detection probability depends on the data, as this is the input to a detection pipeline when assigning a FAR. The normalization in the denominator also corresponds to the fraction of detectable events expected from the population given by $\Lambda$ \citep{Farr19}. We make the following definition of the denominator:
\begin{equation}\label{mu}
    \mu(\Lambda) \equiv \int\int dd_i\, d\theta_i \, p(d_i|\theta_i)\, p(\theta_i|\Lambda)\, p_{\rm det}(\theta_i, d_i)
\end{equation}

The total likelihood comes from considering the probability of the entire dataset $\{d_i\}$ of $N_d$ detections (where the i$th$ event is detected if $p_{\rm det}$ is 1), given a population with parameters $\Lambda$ that predicts $N$ total events, $N \mu \equiv K(\Lambda)$ of which are expected to be detected. The total likelihood is just the product of the contributions from all the detected events, and the likelihood of detecting $N_d$ events, considering the realization of $N_d$ comes from a Poisson distribution with expected value $K$:

\begin{equation}
    p(\{d\}|\Lambda, K) = p(N_d|K(\Lambda)) \prod^{N_d}_{i} p(d_i|\Lambda) \propto K(\Lambda)^{N_d} e^{-K(\Lambda)} \mu(\Lambda)^{-N_d} \prod^{N_d}_i \int d\theta_i p(d_i|\theta_i) p(\theta_i|\Lambda).
\end{equation}

If we assume a prior of $\pi(K) \propto {1}/{K}$, we can write the posterior over $\Lambda$ and analytically integrate out the distribution over $K$:

\begin{equation}\label{fullposterior}
\begin{split}
p(\Lambda|\{d\}) & \propto \pi(\Lambda) \int dK \frac{K(\Lambda)^{N_d} e^{-K(\Lambda)}}{K} \mu(\Lambda)^{-N_d} \prod^{N_d}_i \int d\theta_i p(d_i|\theta_i) p(\theta_i|\Lambda) \\&= \Gamma(N_d) \pi(\Lambda) \mu(\Lambda)^{-N_d} \prod^{N_d}_i \int d\theta_i p(d_i|\theta_i)p(\theta_i|\Lambda) \\& \propto \pi(\Lambda) \mu(\Lambda)^{-N_d} \prod^{N_d}_i \int d\theta_i p(d_i|\theta_i)p(\theta_i|\Lambda)
\end{split}
\end{equation}

where $\Gamma(N_d)$ does not depend on $\Lambda$, so marginalizing over $K(\Lambda)$ with this choice of $\pi(K)$ allows us to factorize the above equation, without explicitly considering its dependence on the Poisson term. 

In practice, $p(\theta|\Lambda)$ does not need to be normalized, as any prefactors will divide out in Eq.~\ref{pereventlikelihood}. We therefore only need to calculate something proportional to $p(\theta|\Lambda)$. For reasons that will become apparent, we compute $p(\theta|\Lambda)$ in terms of something proportional to $\frac{dN}{d\theta}(\Lambda)$.
We want to define a normalization factor for the population distribution such that:
\begin{equation}
    \frac{1}{\alpha(\Lambda)} \left. m_1 \frac{dN}{dm_1 dq dV dt_s}\right|_{(m_{\rm ref}, q_{\rm ref}, z_{\rm ref})} = 1
\end{equation}
where the differential rate is evaluated at a set of reference parameters.

With the distributions in Sec.~\ref{sec:models} defined in terms of $\frac{dN}{dm}$ and $\frac{dN}{dVdt_s}$ (i.e. source frame merger rate density $\mathcal{R}(z)$), we can compute a normalization factor $\alpha(\Lambda)$:
\begin{equation}\label{normalization}
\begin{split}
    \alpha(\Lambda) = \left. m_1\frac{dN}{dm_1 dq dV dt_s}(\Lambda)\right|_{(m_{\rm ref}, q_{\rm ref}, z_{\rm ref})} = & \left. m_1 \frac{dN}{dm_1} \frac{dN}{dm_2} \frac{dm_2}{dq} \frac{dN}{dVdt_s}\right|_{(m_{\rm ref}, q_{\rm ref}, z_{\rm ref})}   \\
     & = \left. m_{1}^2 \frac{dN}{dm} \frac{dN}{dm} \frac{dN}{dVdt_s}\right|_{(m_{\rm ref}, q_{\rm ref}, z_{\rm ref})}
\end{split}
\end{equation}
Technically, we only know the $dN$ distributions up to a constant. As we will see below, we will only be considering ratios of values that share the same unknown constant, so we are free to leave it out for now.

Instead of computing $p(\theta|\Lambda)$ as $p(\theta|\Lambda) = \frac{1}{N} \frac{dN}{d\theta}(\Lambda)$ exactly, we instead make the following transformation in Equations \ref{pereventlikelihood} and \ref{fullposterior}:

\begin{equation}\label{replace_ptheta}
    p(\theta|\Lambda) \rightarrow \frac{1}{\alpha(\Lambda)}\frac{dN}{d\theta}(\Lambda)
\end{equation}
which is directly proportional to the differential rate and $p(\theta|\Lambda)$. 

For each draw of $\Lambda$, we have the normalization factor $\alpha(\Lambda)$, related to the differential rate at our reference parameters, as defined in Eq.~\ref{normalization}. We outline below how we use this re-expression to construct the rate independent of the likelihood. Note that this change in Eq.~\ref{replace_ptheta} does not affect the likelihood, as it only affects $\theta$-independent prefactors, which factor out of both the numerator and denominator in Eq.~\ref{pereventlikelihood}. 

Given the values of $\alpha(\Lambda)$ we have calculated, we wish to draw new samples of $\alpha(\Lambda)$, given that the number of detections is a Poisson-distributed realization.
Recalling $K \equiv N\mu$, Eq.~\ref{replace_ptheta} means that when we compute the denominator of Eq.~\ref{pereventlikelihood}, we are actually calculating the ratio $\frac{K}{\alpha}$, and not $\mu$. Noting that the K-dependent integrand of Eq.~\ref{fullposterior} is a Gamma-distribution for $K$ with shape parameter $N_d$ and a scale parameter of 1, we can make the identification that $\left<K\right> = N_d$ under this distribution. With $K \equiv N\mu$, we can express the expectation value for  $\alpha$ as $\left<\alpha\right> = \frac{N_d}{K/\alpha}$.\footnote{Since $\left<K\right> = N_d$, it follows that $\sigma^2_{\alpha} = \frac{N_d}{(K/\alpha)^2}$} As a final step in post-processing, we can construct the true underlying distribution for $\alpha$ by drawing samples $\alpha \sim \rm Gamma(\frac{N_d}{K/\alpha}, 1)$. This gives us a distribution for the predicted merger rate at the reference coordinates, given the normalization factor $\alpha$ we computed during the hierarchical inference, assuming this is Poisson-distributed about the true value and assuming a $\frac{1}{K}$ prior. With the distribution of $\alpha$, we can scale $\frac{dN}{d\theta}$ to get the differential merger rate at any set of coordinates $\theta$.

Note that we have written everything in this section in terms of $\theta$ as if it is always the parameters in the population model, suppressing the fact that there will be Jacobians in Eq.~\ref{pereventlikelihood} to transform from these coordinates to those in the detector-frame (or the priors from the single-event analyses).

\section{Accounting for Spin Distribution}\label{app:spins}
In the analysis presented in the body of this work, we assume the (uninformative) parameter estimation priors in the population reweighting. Based on population-level mass-spin and mass-redshift correlations presented in the literature (see, e.g., \cite{Callister21, Biscoveanu22, gwtc3pops}), we may expect the assumed spin distribution can have an effect on our results. However, with the relatively poor spin constraints in the population, we empirically demonstrate that this is likely not the case.

\begin{figure}
    \centering
    \includegraphics[scale=0.4]{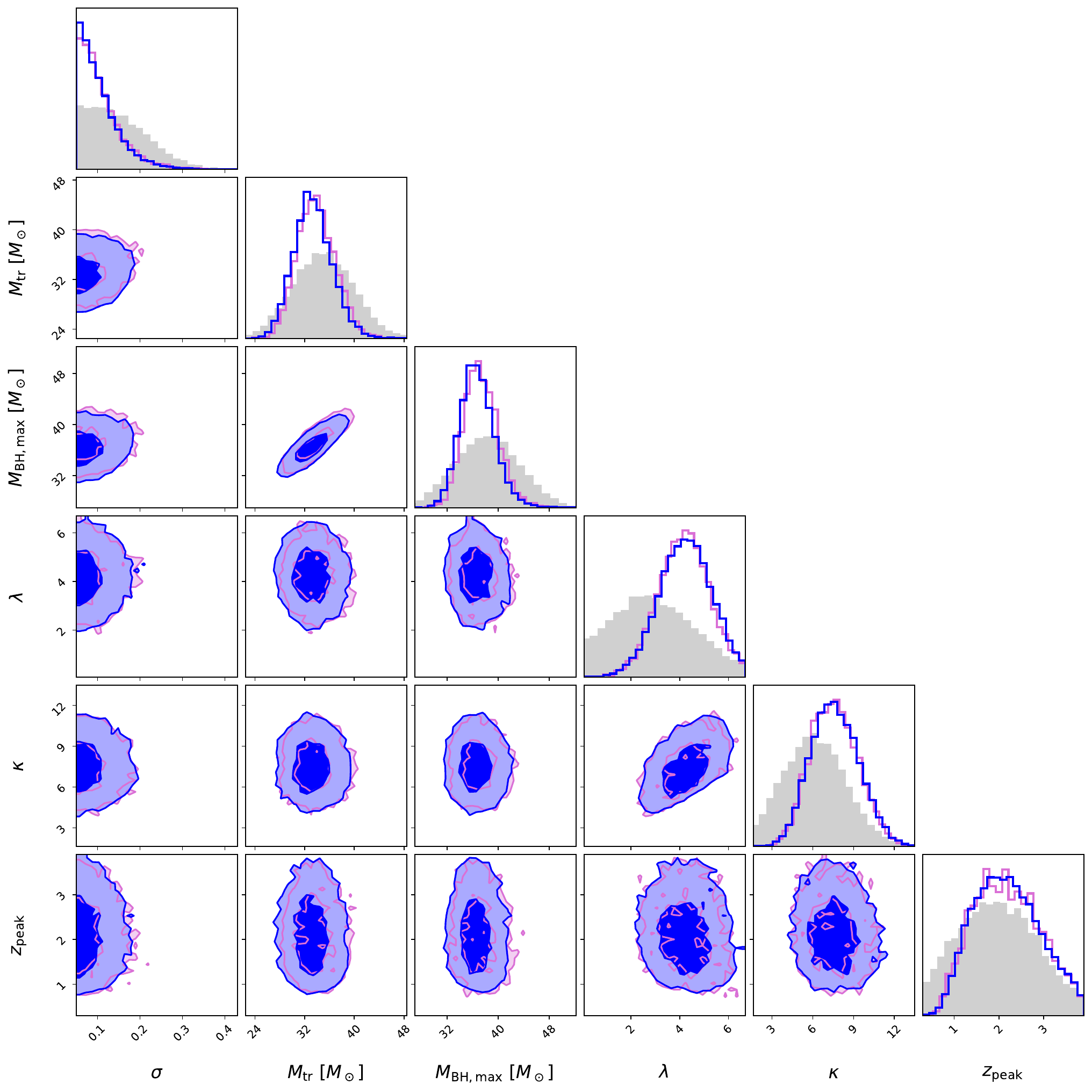}
    \caption{Selected population-level parameters from the evolving mass distribution analysis, without reweighting spins (pink, as presented in Section~\ref{sec:GWTC3evolvingresults}) and with reweighting spins to a fiducial population estimate (blue). The near-identical posteriors show that the spin population assumptions in this work do not cause a bias.}
    \label{fig:corner_spins_reweighted}
\end{figure}

In Figure~\ref{fig:corner_spins_reweighted}, we compare posteriors obtained from our main analysis ignoring spins, with those obtained by reweighting the posterior samples from each event and the sensitivity injections to a fiducial spin distribution. For this fiducial spin distribution, the spin magnitudes are from a half-Gaussian centered at $a = 0$ with a standard deviation of 0.3, meant to model the preferentially-small spin mangitudes inferred in \cite{gwtc3pops}. For the contribution aligned projection of the spin tilt angle ($\cos\theta$), we use the mixture model introduced in \cite{Talbot17}, with an aligned-spin fraction of $\xi = 0.8$ and an aligned-spin spread of $\sigma_t = 1.9$, consistent with the results reported in \cite{gwtc3pops} (see references for definitions of these model parameters). We find that reweighting to this spin distribution has a negligible effect on our inferred population.

\section{Details on PPISN Interpretation}\label{app:ppisndetails}

The physical reason the the anticorrelation between $\rm ^{12}C(\alpha, \gamma)^{16}O$ and $M_{\rm BH,max}$ is that during contraction of the stellar core, hydrostatic equilibrium can be maintained by convective carbon burning. Higher $\rm ^{12}C(\alpha, \gamma)^{16}O$ reaction rates lead to cores of lower carbon fractions, $X_C$. When the core gets hot enough to produce electron-positron pairs, the equation of state softens, leading to a contraction. With little carbon present to provide convective-driven pressure to stabilize the star, contraction can continue until it drives thermonuclear ignition of oxygen. This explosive process leads to an outward-moving shock, removing mass from the star when the shock reaches the surface with enough velocity. Once this shock breaks through the surface of the star, contraction begins again. This sequence of pulsations continues until oxygen in the core is depleted, core elements burn through the normal pre-SN process, and the star undergoes normal core collapse. If the $^{12}C(\alpha, \gamma)^{16}O$ reaction rate is lower, then relatively more carbon is present and able to burn convectively, counteracting the contractions in a stable manner. As the carbon fraction gets higher, the star is able to remain stable against pair-production contractions and stably burn through the core oxygen \citep{Farmer20, Woosley17, Woosley19}

The location for the onset of Pair Instability Supernova (PISN) is highly sensitive to the $\rm ^{12}C(\alpha, \gamma)^{16}O$ reaction rate. With the core temperature strongly increasing with stellar mass, there exists a core mass at which the softening of the equation of state is too extreme to be resisted by available sources of outward pressure. Since stable outward pressure support at this stage is largely provided by shell carbon-burning, the lower carbon fraction, $X_C$, makes it now easier for a given contraction to compress and fully ignite the oxygen core, driving a subsequent pulsation so powerful that further pair production in the core cannot re-soften the equation of state fast enough to return it to a contraction phase. This is basically equivalent to a single pulsation during the pulsational pair-instability process blowing away the total mass of the star \citep{Woosley17, Woosley19, Woosley21}. A lower carbon fraction results in this full disruption of the star (PISN) occurring at lower masses, controlling where the $\MI-M_{\rm BH}$ map decays to zero after $M_{\rm BH,max}$.

\bibliography{bibliography}
\end{document}